\documentclass[final,3p,times]{elsarticle}
\usepackage{amssymb}
\usepackage{amsmath}
\usepackage{hyperref}
\usepackage[hyperref]{xcolor}

\hypersetup{
  colorlinks=true,
  linkcolor = red,
  urlcolor  = magenta,
  citecolor = blue,
  anchorcolor = blue
}

\begin{document}

\hypersetup{
  linkcolor = red,
  urlcolor  = magenta,
  citecolor = blue,
  anchorcolor = blue
}

\newcommand{\rev}[1]{{\color{purple}{#1}}}

\begin{frontmatter}

\title{On Evidence for Elastic Interactions in the Dark Sector} 

\author[1,2]{Jose Beltr\'an Jim\'enez} 
\ead{jose.beltran@usal.es}
\author[2,3]{Dario Bettoni}%}\fnref{fn2}} 
\ead{dbet@unileon.es}
\author[1,2,5]{David Figueruelo}%\fnref{fn1,fn3}} 
\ead{davidfiguer@usal.es}
\author[4,2]{Florencia A. Teppa Pannia}%\fnref{fn1,fn3}} 
\ead{fa.teppa.pannia@upm.es}
%\fntext[fn1]{This is the first author footnote.} \fntext[fn2]{Another author footnote, this is a very longfootnote and it should be a really long footnote. But this footnote is not yet sufficiently long enough to make two lines of footnote text.}
%\fntext[fn3]{Yet another author footnote.}

\affiliation[1]{organization={Departamento de F\'isica Fundamental, Universidad de Salamanca},
%%           addressline={},
           city={Salamanca},
            postcode={E-37008},
%%           state={},
             country={Spain}}

\affiliation[2]{organization={Instituto Universitario de F\'isica Fundamental y Matem\'aticas~(IUFFyM), Universidad de Salamanca},
%%           addressline={},
           city={Salamanca},
            postcode={E-37008},
%%           state={},
             country={Spain}}             

\affiliation[3]{organization={Departamento de Matem\'aticas, Universidad de Le\'on},
           addressline={Escuela de Ingenier\'ias Industrial, Inform\'atica y Aeroespacial Campus de Vegazana, s/n},
           city={Le\'on},
            postcode={24071},
%%           state={},
             country={Spain}}
             
\affiliation[4]{organization={Departamento de Matem{\'a}tica Aplicada a la Ingenier{\'i}a Industrial, Universidad 
 Polit{\'e}cnica de Madrid},
%%           addressline={},
           city={Madrid},
            postcode={E-28006},
%%           state={},
             country={Spain}}

\affiliation[5]{organization={Centre for Space Research, North-West University},
%%           addressline={},
           city={Potchefstroom},
            postcode={2520},
%%           state={},
             country={South Africa}}

%% Abstract
\begin{abstract}
The increasing quality of cosmological data has revealed some tensions that could be signalling the necessity of incorporating new physics into our cosmological model. One particularly intriguing possibility is the existence of elastic interactions between dark matter and dark energy. Not only do these interactions provide a natural mechanism to relief cosmological tensions, but there is also compelling observational evidence for them, to the extent that a detection could even be claimed. We review the potential of these scenarios in relation to the cosmological tensions and discuss distinctive signatures that can be probed with future data, thus providing  a smoking gun for these interactions.
\end{abstract}



%% Keywords

\end{frontmatter}


\section{Introduction}
\label{sec:intro}

Traditionally, cosmology has served as fertile ground for theoretical speculation, only limited by the bounds of imagination. This speculative nature of cosmology was largely due to a scarcity of data guiding theoreticians. However, this scenario has undergone a dramatic transformation in recent decades, owing to tremendous technological advancements that have enabled the construction of better instruments for observing the cosmos. The observations gathered with these instruments have delineated the realms of permissible speculation. Paradigmatic examples of how data have shaped our comprehension of the universe include the anisotropies in the Cosmic Microwave background (CMB) by satellites such as WMAP \cite{WMAP:2012nax} and Planck \cite{Aghanim:2018eyx}, as well as the mapping of Large Scale Structures (LSS) of the universe through surveys like SDSS  \cite{eBOSS:2020yzd}, DES \cite{DES:2021bvc,DES:2021vln,DES:2021wwk}, KiDS \cite{Kuijken:2015vca,KiDS:2020suj}, HSC \cite{Dalal:2023olq,Li:2023tui} or DESI \cite{DESI:2023ytc}. These observations have not only helped establishing $\Lambda$CDM as the standard model of cosmology, but they have also provided robust evidence for the existence of a dark sector comprised of dark matter and dark energy.

While cosmological observations have indeed tempered the speculative nature of cosmological models, it is also a common occurrence in physics that an increase in the precision of the data comes with the emergence of anomalies that potentially challenge our physical understanding. Initially, these anomalies may not appear particularly significant, and some may even dissipate with additional data or more refined analysis. However, in other instances, seemingly innocuous anomalies can ultimately pave the way for groundbreaking discoveries, potentially precipitating a paradigm shift. The appearance of some tensions in cosmological data \cite{Perivolaropoulos:2021jda,Abdalla:2022yfr}, being particularly pressing the $H_0$ and the $\sigma_8$ tensions although the more recent DESI data has contributed a potential new "$w_0-w_a$" tension \cite{DESI:2024uvr,DESI:2024mwx}, make it plausible that cosmology is presently experiencing one of these revolutionary periods, thus being a unique opportunity to discover novel phenomena. Given the elusive character of the dark sector, it is particularly alluring to seek for new physics associated to dark matter and/or dark energy. This brings us back to theoretical speculation, but this time being guided by very restrictive data.

In this brief review, we will be concerned with the $\sigma_8$ tension. As we reach the completion of Stage-III LSS experiments that have provided data with remarkable precision, we witness how inferences from different sources appear to indicate that the clustering of dark matter seems to be systematically weaker than the one expected from extrapolating CMB measurements within $\Lambda$CDM. The amount of clustering can be conveniently encoded in the parameter $\sigma_8$ (or its relative\footnote{Although we will refer to the $\sigma_8$ and $S_8$ tensions as somewhat interchangeable  terminology referring to the same cosmological conundrum apparently indicating a clustering deficit, the precise relation between both quantities depends on $\Omega_{\text m}$ so, strictly speaking, they are only equivalent if $\Omega_{\text m}$ is known. See e.g. \cite{Sakr:2023hrl,Poulin:2024ken,Pedrotti:2024kpn} for relevant discussions related to the role of $\Omega_{\text m}$ in the cosmological tensions.} $S_8$) that gives a measure of the amplitude of matter fluctuations in the present universe and the values reported by low-redshift probes show a trend to be systematically smaller than the value measured by the Planck satellite \cite{Perivolaropoulos:2021jda,Abdalla:2022yfr}. These observations might signal that matter clusters less than expected between matter-radiation decoupling and today. Interestingly, the Atacama Cosmology Telescope (ACT) \cite{ACT:2023kun} has recently reported values of $\sigma_8$ from CMB lensing measurements that are compatible with Planck. The relevance of this measurement is that ACT probes linear scales for redshifts $z\simeq 0.5-5$ with its sensitivity peaking at $z\sim 2$, thus suggesting that, if matter indeed clusters less than expected, the mechanism responsible for such a clustering deficit should be at work in the very late universe $z\lesssim 2$.

Guided by the observational evidence, it is natural to speculate the existence of a mechanism in the dark sector preventing the matter clustering at low redshift. However, the mechanism must be judiciously constructed since it should explain the reduced clustering while still accommodating the data explained by the standard model. After some reflections, we come to the conclusion that pressure is a natural agent to prevent matter clustering, so it is appealing to employ it for our purpose. As a matter of fact, the pressure of photons is precisely what prevents the clustering of baryons before decoupling, when they interact via Thomson scattering. This motivates to use the dark energy pressure to prevent the clustering of dark matter, a scenario that can be realised by introducing new interactions in the dark sector.\footnote{The history of interactions in the dark sector is very long and a very extensive literature exists with a plethora of different interacting models. We refer to \cite{Wang:2024vmw} for a comprehensive review of interacting models.} This interaction should be sensibly manufactured though since we do not want to spoil all the successes of the standard model. In particular, we would like to leave the background cosmology oblivious to the mechanism so we completely disentangle the effects on the clustering from the cosmological expansion. Seeking motivation from pre-decoupling physics again, we can consider an elastic interaction between dark matter and dark energy, i.e., an interaction without energy exchange (at least at first order in cosmological perturbations). This short review is devoted to exploring these scenarios and to put forward their striking nature. 

%%%%%%%%%%%%%%%%%%%%%%%%%%%%%%%%%%%%
\section{A proxy for dark elastic interactions}
\label{sec:interactions}

There are different manners to realise the idea of having elastic interactions in the dark sector. One option is to assume the presence of elastic scatterings between dark matter and dark energy {\it particles} \cite{Simpson:2010vh}. However, the corresponding interaction rate decreases with the expansion so it is only important at high redshift and it dies away as the universe expands. Thus, it does not fully comply with our requirements. More generally, one can study the possible interactions in the dark sector between dark matter characterised as a fluid and dark energy described in terms of a scalar field. This was done in a systematic way in e.g. \cite{Pourtsidou:2013nha,Skordis:2015yra} where the possible interactions were classified into three types, one of which corresponds to pure momentum transfer between the dark components. Different versions of this scenario as well as extensions have been studied in e.g. \cite{Koivisto:2015qua,Kase:2019mox,Chamings:2019kcl,Amendola:2020ldb,Linton:2021cgd,ManciniSpurio:2021jvx} and similar scenarios where dark energy is described by a vector rather than a scalar field have been considered in \cite{Nakamura:2019phn,DeFelice:2020icf,Cardona:2023gzq,Pookkillath:2024ycd}. An Effective Field Theory approach was employed in \cite{Gleyzes:2015pma} to describe certain interacting scenarios with dark energy and modified gravity. A more phenomenological scenario for elastic interactions at the level of the conservation equations was introduced in \cite{Asghari:2019qld} and a Lagrangian formulation of similar models was developed in \cite{BeltranJimenez:2020qdu}. Barring some particularities of each model, the scenarios featuring an elastic interaction in the dark sector share some common interesting properties. In particular, as anticipated, they can generically take advantage of the dark energy pressure to produce a weaker growth of structures, thus providing compelling scenarios for alleviating the $\sigma_8$ tension. It is worth mentioning that models with an elastic coupling between dark energy and baryons have also been explored in e.g. \cite{Vagnozzi:2019kvw,BeltranJimenez:2020iyx}. It was shown in \cite{BeltranJimenez:2020iyx} that a reduction of the matter clustering can also take place, although less prominently than in the dark matter case due to the smaller abundance of baryons. Elastic interactions of dark matter with photons \cite{Wilkinson:2013kia,Stadler:2018jin,Kumar:2018yhh} and neutrinos \cite{Stadler:2019dii} have also been explored. However, in all these cases the interactions are more tightly constrained because they give rise to novel scattering channels that would affect e.g. measurements from collider experiments.

For the sake of concreteness, we shall focus here on the scenario introduced in \cite{Asghari:2019qld} and subsequently explored in \cite{Figueruelo:2021elm,BeltranJimenez:2021wbq,Cardona:2022mdq,Poulin:2022sgp}. The reason for choosing this model as a proxy for the class of cosmologies with dark elastic interactions is its simplicity, as we will see in the following, as well as its ability to leave the background evolution completely unaffected.\footnote{In the pure momentum transfer models of \cite{Pourtsidou:2013nha} for instance, one still needs to tackle the background evolution of the scalar field.} The main idea is to modify the usual energy-momentum conservation equations by adding an interaction proportional to the relative velocity of the dark components. Since all the components share a common large scale rest frame, as dictated by the Cosmological Principle,  the interaction is innocuous on very large scales. On small scales however, the existence of peculiar velocities will make the interaction operative. Since dark energy becomes important at late times and peculiar velocities also appear as structures form, the interacting epoch will naturally commence at late times, although this will also depend on the cosmological evolution of the interaction rate. We will then consider a modification of the conservation equations in the dark sector described by
\begin{equation}
\nabla_\nu T_{\text{dm}}^{\mu\nu}=\alpha\Big(u_{\text{dm}}^\mu-u_{\text{de}}^\mu \Big)\,,\qquad
\nabla_\nu T_{\text{de}}^{\mu\nu}=-\alpha\Big(u_{\text{dm}}^\mu-u_{\text{de}}^\mu \Big)\,,
\label{eq:interactingT}
\end{equation}
where $\alpha$ is some constant parameter that measures the strength of the interaction and $u^\mu_a$ is the 4-velocity of the corresponding fluid. For dark matter we will utilise the usual assumptions of being cold, so its pressure and sound speed are nearly vanishing, while for dark energy we will assume a constant equation of state parameter $w$ and some sound speed $c_{\text{de}}^2$, which we will eventually set to 1 in order to have pressure perturbations able to efficiently prevent the clustering. It is apparent from Eqs.~\eqref{eq:interactingT} that the interaction only operates when there is a relative motion between the dark components, i.e., on sub-Hubble scales, while the background evolution is not affected by the interaction. The perturbed dark sector described by the density contrasts $\delta_j=\delta\rho_j/\rho_j$ and the peculiar velocities $\theta_j= i \vec{k}\cdot\vec{v}_j$, where $j$ stands for dark matter or dark energy, is governed, in Fourier space and assuming no anisotropic stresses, by the equations\footnote{From now on, the parameter $\alpha$ will be measured in its natural units $\alpha\to \frac{\alpha}{\rho_c H_0}$, with $\rho_c$ and $H_0$ the critical density and Hubble factor today.}
\begin{eqnarray}
\label{eq:deltaDM}
\delta_{\rm dm}' &=& 
-\theta_{\rm dm}+h_A\,,\\
\delta_{\rm de}'&=&-3 \mathcal{H}\left( c_{\rm de}^2-w\right) \delta_{\rm de} +(1+w)h_A  -(1+w)\left(1+9 \mathcal{H}^2
\frac{c_{\rm de}^2-w}{k^2}\right)
\theta_{\rm de}\;,  \\
\label{eq:thetaDM}
\theta_{\rm dm}'&=&-\mathcal{H} \theta_{\rm dm} + k^2\Phi  + \Gamma_\alpha\Big(\theta_{\rm de}-\theta_{\rm dm}\Big)\;, \\
\label{eq:thetaDE}
\theta_{\rm de}' &=&\left(3c_{\rm de}^2-1\right) \mathcal{H}\theta_{\rm de}+k^2\Phi +\frac{c_{\rm de}^2k^2}{1+w}\delta_{\rm de} 
 -\Gamma_\alpha R\Big(\theta_{\rm de}-\theta_{\rm dm}\Big)\,,
\end{eqnarray}
where $\mathcal{H}$ is the Hubble function in conformal time and we have defined the following convenient quantities
\begin{equation}
\Gamma_\alpha\equiv \alpha \frac{a^4}{\Omega_{\rm dm}}\,,\qquad R \equiv
\frac{\Omega_{\rm dm}}{(1+w)\Omega_{\rm de}}a^{3w}\,. 
\end{equation}
The above perturbation equations are valid in both Newtonian and synchronous gauge with the metric perturbation $h_A$ taking the values, in the notation of \cite{Ma:1995ey}, $h_{\text{Newtonian}}=3\Phi'$ and $h_{\text{synchronous}}=-\frac12 h'$ and the understanding that the term $k^2\Phi$ in \eqref{eq:thetaDM} and \eqref{eq:thetaDE} should be removed in the synchronous gauge. Our discussions on the interacting scenario below will be mainly referred to the Newtonian gauge. However, an important property of the interaction is its gauge invariance (at linear order) because velocity perturbations change with the time derivative of the spatial gauge parameter, so the gauge dependence disappears when we take the difference of any two velocity perturbations. In other words, although we need a reference frame to define a velocity, the relative velocities do not depend on the frame choice. Thus, the interaction term can be defined and analysed in any frame and it is not affected by possible superfluous gauge modes as those present in e.g. the synchronous gauge.  The function $\Gamma_\alpha$ describes the interaction rate, while $R$ is the relative abundance of dark matter to dark energy. As advertised, only the Euler equations are modified by the interaction and the relation between the density contrast and the peculiar velocities is the same as in the standard scenario. This feature is important because it allows to infer the density perturbation from the velocity field as in the standard model, despite having an interacting dark sector. In more general interacting scenarios, this is not the case and additional care must be taken when confronting the model predictions to observations. Let us notice the formal resemblance of the Euler equations \eqref{eq:thetaDM} and \eqref{eq:thetaDE} with the equations governing the baryon-photon plasma before decoupling (see e.g. \cite{weinberg2008cosmology}) where Thomson scattering generates an analogous term proportional to the relative velocities.\footnote{This analogy has motivated to call it sometimes covariantized dark Thomson interaction. We will refer to it as $\alpha$CDM for obvious reasons.} There are some important differences however. While Thomson scattering gives rise to an interaction rate that decreases as the universe expands, $\Gamma_\alpha$ grows as $\propto a^4$ and this ensures that the universe will enter a {\it dark-coupling epoch} at late times as it expands, rather than undergoing a decoupling phase. This is a crucial difference of this scenario with the elastic dark scattering considered in \cite{Simpson:2010vh}  which does exit rather than entering a dark coupled epoch.\footnote{This is also the reason why no relevant observational effects were found in the baryons-dark energy elastic scattering studied in \cite{Vagnozzi:2019kvw} while important effects were observed in \cite{BeltranJimenez:2020iyx}.} Some observational constraints on this model have been obtained in \cite{Kumar:2017bpv}.

\begin{figure}[t]
\includegraphics[width=0.49\textwidth]{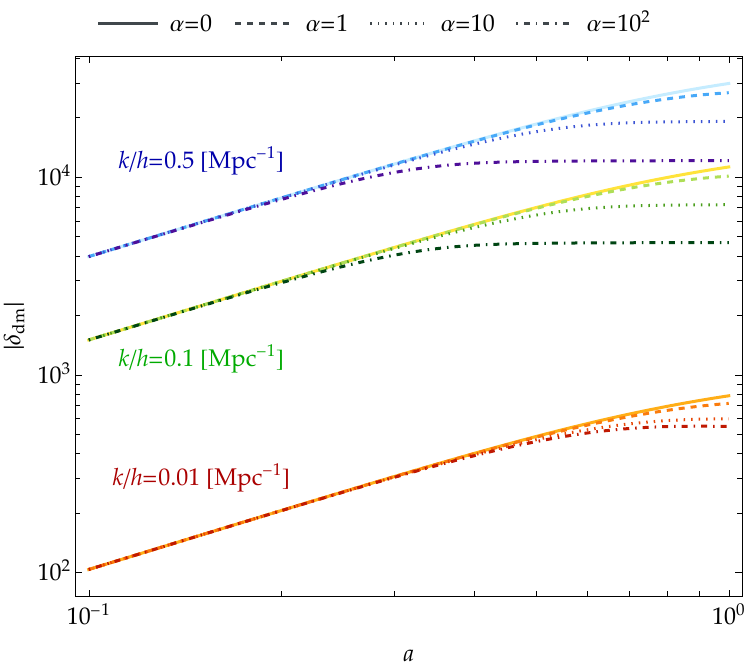} \hspace{0.1cm}
\includegraphics[width=0.49\textwidth]{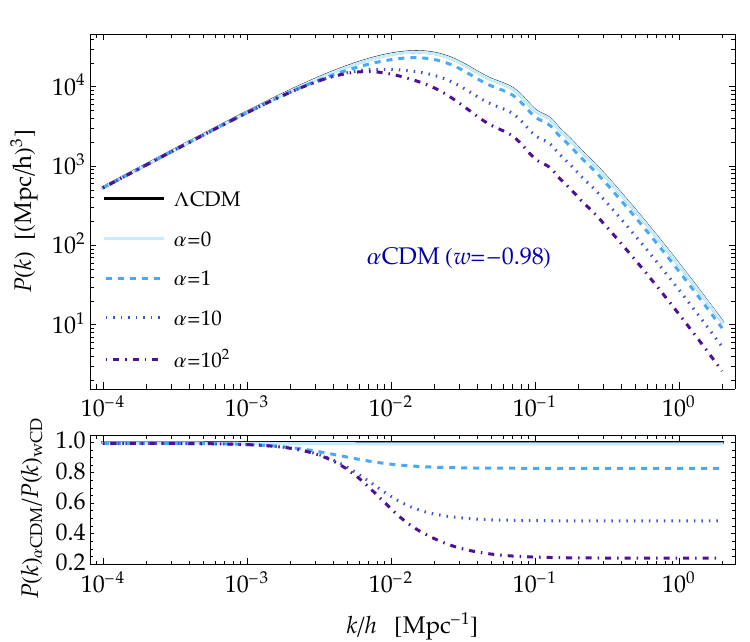}
\caption{This Figure shows the effects on the growth of structures (left) and the matter power spectrum (right) for different values of the interaction parameter $\alpha$ and in a $w$CDM model with $w=-0.98$ as computed from a modified version of CAMB. We corroborate the usual growth $\delta_{\text{dm}}\propto a$ until the beginning of the dark-coupling epoch, which occurs earlier for higher values of $\alpha$. Modes that have entered the horizon before the dark-coupling epoch freeze their growth (blue and green) while modes that enter later (orange) keep growing until they cross the horizon. As explained in the main text, this causes a suppression in the matter power spectrum as shown in the right panels. The suppression is $k$-independent for scales smaller than the horizon at the dark-coupling time, while larger scales (but smaller than the horizon today) undergo a partial suppression.
} 
\label{Fig:Pk}
\end{figure}

The dark matter Euler equation \eqref{eq:thetaDM} tells us that the interaction will affect the dark matter clustering in a regime where $\Gamma_\alpha\gtrsim\mathcal{H}$, which marks the onset of the dark-coupling epoch. Once the interaction is fully operative, the peculiar velocities of dark matter will decrease as compared to the non-interacting scenario due to the drag exerted by the dark energy pressure. This drag will then be the responsible for the suppression of the clustering via the continuity equation \eqref{eq:deltaDM}. As a matter of fact, the drag is so efficient that the dark matter density contrast completely ceases its growth and it remains constant throughout the dark-coupling epoch (see left panel in Figure~\ref{Fig:Pk}). This means that the density contrast freezes with the amplitude it has when it enters the dark-coupling epoch and it remains with that amplitude until today, thus originating a suppression in the clustering with respect to $\Lambda$CDM. 

Since the density contrast evolves as $\delta_{\text{dm}}\propto a$ in $\Lambda$CDM, we can easily estimate that the interaction will induce a suppression by a factor $\frac{a_\alpha}{a_0}$, where $a_\alpha$ is the scale factor at the beginning of the dark-coupling epoch and $a_0$ the scale factor today. A remarkable property of this suppression is that it is the same for all the $k$-modes since it is determined by the time when the dark-coupling epoch commences. This is true for modes that are already sub-Hubble at the beginning of the dark-coupling epoch so that they have always felt the interaction. However, modes that cross the horizon in the dark coupling epoch, only suffer a partial suppression because they are oblivious to the interaction until they enter the horizon. If the horizon crossing occurs at $a_*$, then the corresponding density contrast is only suppressed by a factor $a_*/a_0$. This suppression is smaller for larger scales because their horizon crossing occurs later and, therefore, they suffer a smaller suppression until we reach the limiting case of the modes that have not crossed the horizon today, for which there is no suppression.

The discussed phenomenology fully explains the effects on the matter power spectrum that we see in the right panel of Figure~\ref{Fig:Pk}. In addition to the commented suppression, a very distinctive feature that we can also observe is the shift in the peak of the power spectrum caused by the interaction. The importance of this effect resides in the fact that the matter power spectrum turnover is usually fixed by the horizon scale at equality time, when matter and radiation are equally abundant, so it is solely determined by background quantities. In the interacting scenario, the shift is provoked by the evolution of the perturbations that are in turn only modified at late times. Thus, while the peak is related to the evolution of the perturbations in the early universe and fixed by background parameters in standard scenarios, the interaction is able to produce a shift via a parameter that exclusively appears at the level of the perturbations and from a mechanism that operates at late times. 

The mechanism by which the dark elastic interaction induces a suppression of the clustering, thus potentially addressing the $\sigma_8$ tension, complies with all the properties we asked for. It is a mechanism that leaves the background intact because the parameter $\alpha$ does not play any role for it. It works at late times so the clustering process is identical to the standard model throughout most of the universe history, as required by the recent results of ACT \cite{ACT:2023kun} suggesting that the lack of clustering should come from $z\lesssim 2$. Finally, an additional advantage lies in the minimalist nature of the model, with all the effects being governed by one single parameter $\alpha$. This very single parameter controls both the strength of the interaction by setting the amount of momentum transfer and, at the same time, the scale where the interaction becomes efficient. This simplicity attributes a certain rigidity to the predictions of the model, which in turn is a positive feature because it permits to rule it out, and grants it a Bayesian reward.

%%%%%%%%%%%%%%%%%%%%%%%%%%%%%%%%%%%%
\section{Mitigating a tension...}
\label{sec:s8tension}

The scenarios featuring an elastic interaction in the dark sector certainly possess the desired properties to be promising candidates to alleviate the $\sigma_8$ tension. However, it is imperative to properly address their true potential and to unveil to what extent the tension can actually be relieved. This question is properly answered by confronting the model to data. Adopting a conservative approach, we first consider a baseline dataset conformed by Planck2018, type Ia supernovae and Baryon Acoustic Oscillations data and through an MCMC analysis we obtain the corresponding constraints.

\begin{figure}[!t]
\includegraphics[width=0.49\textwidth]{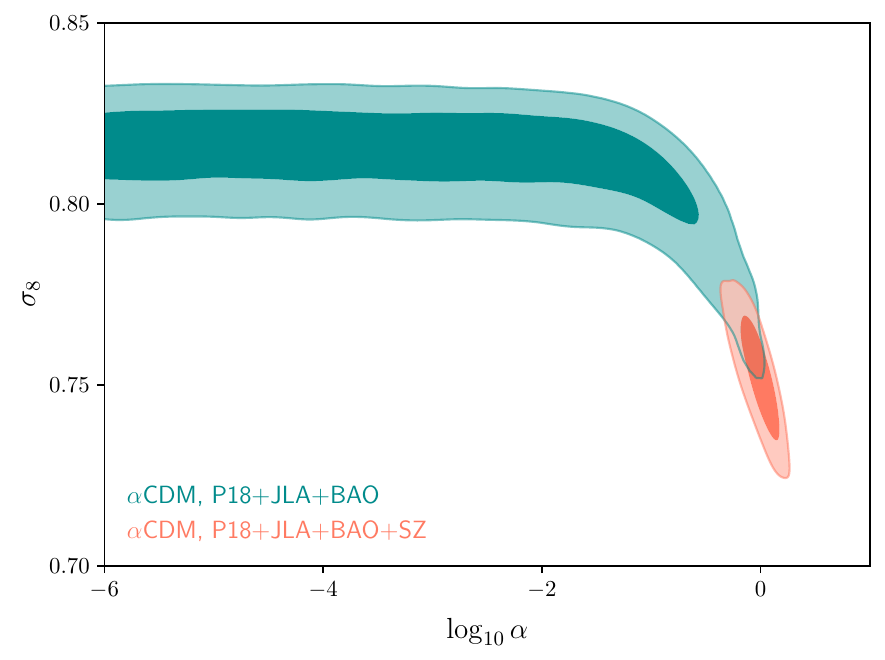} \hspace{0.1cm}
\includegraphics[width=0.49\textwidth]{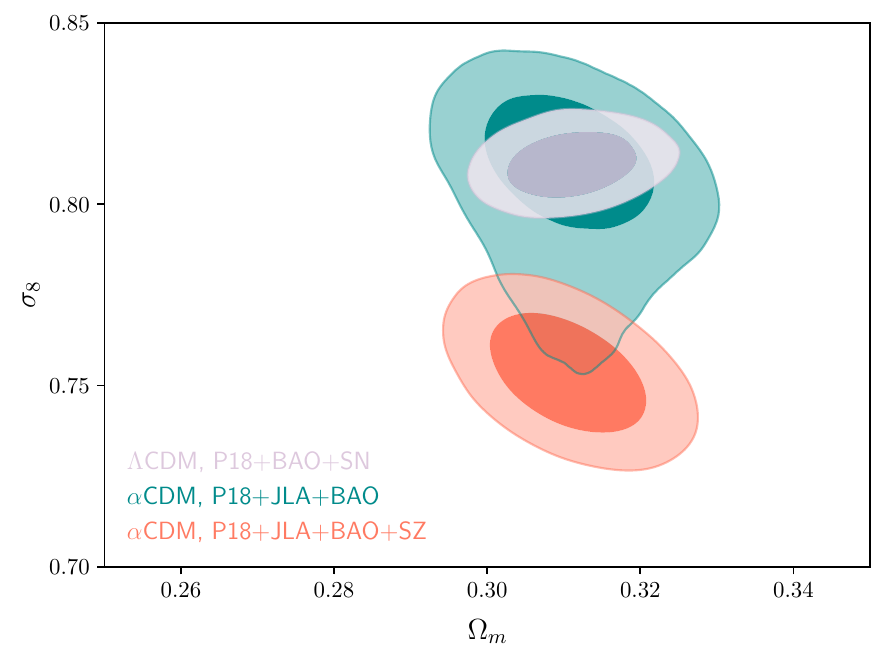}
\caption{The left panel shows the constraints in the $\sigma_8-\log_{10}\alpha$ plane for $\alpha$CDM, while the right panel shows the $\sigma_8-\Omega_m$ plane for $\alpha$CDM and $\Lambda$CDM. In both cases we use the baseline dataset (green) and including a Gaussian prior on $S_8$ (red) from the Planck 2015 Sunyaev–Zeldovich cluster counts as considered in \cite{Figueruelo:2021elm}. We can observe how the interaction allows substantially lower values of $\sigma_8$ than $\Lambda$CDM thus easing the tension with low-redshift probes. Remarkably, adding the $S_8$ data clearly selects a non-vanishing value of $\alpha$.}
\label{Fig:S8-vs-alpha}
\end{figure}

In Figure~\ref{Fig:S8-vs-alpha} we show some relevant marginalised planes from the results of the MCMC analysis as obtained in \cite{Figueruelo:2021elm}, where we refer for further details on the methodology and the employed datasets. As expected, the interaction permits to access a region with lower values of $\sigma_8$. For very small values of $\alpha$, the results for $\sigma_8$ recover the $\Lambda$CDM results and the interaction is irrelevant. However, for values of $\alpha\sim 1$, the contour bends towards lower values of $\sigma_8$, indicating that the interaction is doing its job of suppressing the clustering. This means that the interaction can indeed ease the tension of Planck2018 data with LSS measurements from low redshift. This can be more clearly seen in the $\sigma_8-\Omega_m$ plane shown in the right panel of Figure~\ref{Fig:S8-vs-alpha} where we see how the interaction permits to access significantly lower values of $\sigma_8$ than in $\Lambda$CDM. Furthermore, we also corroborate that the lowering of $\sigma_8$ is fully driven by $\alpha$ and $\Omega_m$ is not affected at all, thus showing how the interaction effects are disentangled from the background parameters. This could be relevant for the results of \cite{Garcia-Garcia:2024gzy} where a combined analysis of weak lensing from DES, KiDS-1000 and HSC-DR1 including small scales is performed. It is found that the $\sigma_8$ tension can be somewhat reduced, but a $\sim 3\sigma$ tension in the value of $\Omega_{\text{m}}$ emerges. This could be due to the need for less matter to reduce the clustering in the standard model. Since the interaction can handle the clustering suppression by itself without affecting $\Omega_{\text{m}}$, a better performance can be expected.

\begin{figure}[t]
\includegraphics[width=0.49\textwidth]{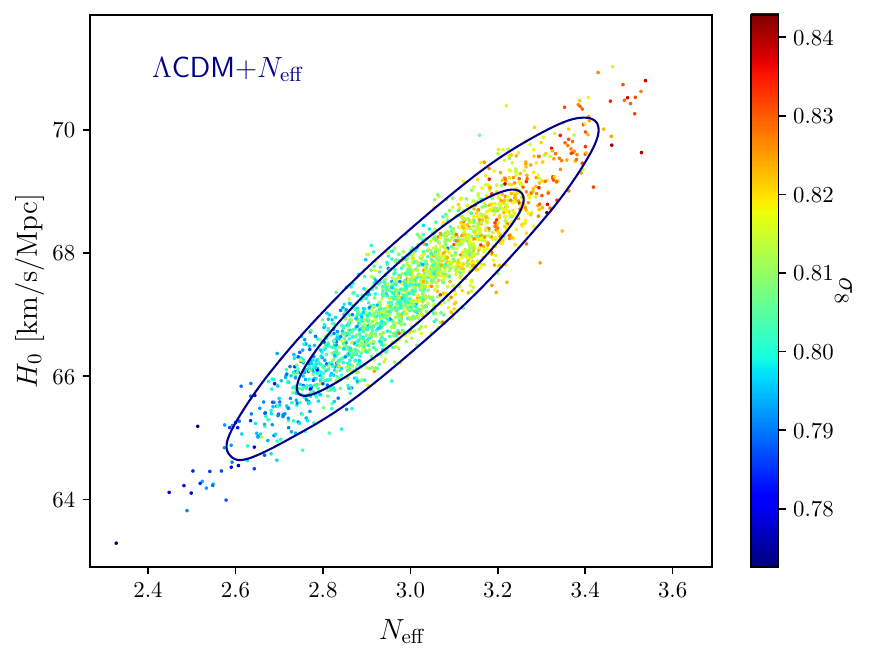} \hspace{0.1cm}
\includegraphics[width=0.49\textwidth]{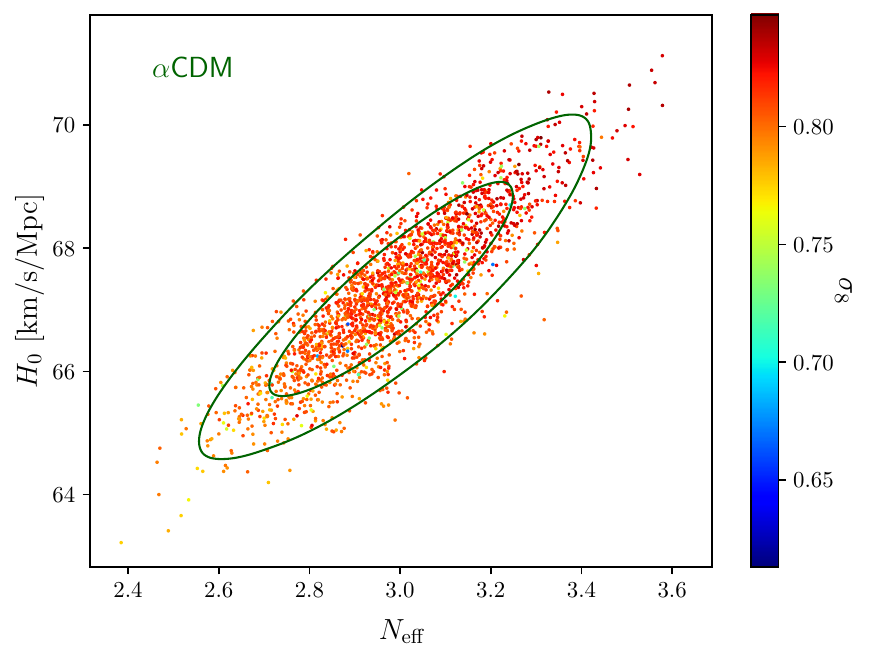}
\caption{Constraints in the $H_0-N_{\text{eff}}$ plane for $\Lambda$CDM (left) and $\alpha$CDM (right) as obtained in \cite{BeltranJimenez:2021wbq}. These plots clearly illustrate how higher values of $H_0$ that would relieve the tension with SH0ES data occur in the region with higher values of $\sigma_8$, thus worsening the $\sigma_8$ tension. On the other hand, we see how this correlation between $H_0$ and $\sigma_8$ disappears in the presence of the interaction.}
\label{Fig:H0Neff}
\end{figure}

The absence of correlations between the interaction parameter $\alpha$ and background quantities acquires especial relevance for the case of $H_0$. This is important because many cosmological scenarios exhibit an anti-correlation between both the $H_0$ and the $\sigma_8$ tensions so that alleviating one of them exacerbates the other (see e.g. \cite{Jedamzik:2020zmd,Poulin:2024ken,Pedrotti:2024kpn}).\footnote{It has even been suggested that the $\sigma_8$ tension might be an $H_0$ tension in disguise \cite{Sanchez:2020vvb,Garcia-Garcia:2024gzy}.} One typical example for this effect occurs when allowing for additional relativistic species as an attempt to solve the $H_0$ tension. This extra radiation does alleviate to some extent the $H_0$ tension, but this requires to go to regions in the parameter space where $\sigma_8$ is even larger, thus worsening the $\sigma_8$ tension (see left panel in Figure~\ref{Fig:H0Neff}). In the presence of the interaction, this is no longer the case since the interaction operates at low redshift to reduce the clustering without affecting the background cosmology, so the alleviation of $H_0$ is untied from the value of $\sigma_8$, as Figure~\ref{Fig:H0Neff} clearly shows.

%%%%%%%%%%%%%%%%%%%%%%%%%%%%%%%%%%%%
\section{... And evidence for a detection?}
\label{sec:s8data}

The results from the confrontation to the baseline datasets confirm the potential of the dark elastic interaction to alleviate the $\sigma_8$ tension. Within these scenarios, it is more consistent to add data from low redshift probes that exhibit a certain tension with Planck2018 data for $\Lambda$CDM. The tension is usually encoded into the parameter\footnote{Another common definition is $S_8\equiv\sigma_8\left(\frac{\Omega_{\text m}}{0.27}\right)^{0.3}$.}
\begin{equation}
S_8\equiv\sigma_8\sqrt{\frac{\Omega_{\text m}}{0.3}}\,,
\end{equation}
which measures the amplitude of the clustering taking into account the correlation with $\Omega_{\text{m}}$. The value reported by Planck2018 is $S_8 = 0.832 \pm 0.013$ \cite{Aghanim:2018eyx} and it is in agreement with other CMB measurements from WMAP \cite{WMAP:2012nax} and ACT \cite{ACT:2023kun}. The value of $S_8$ can alternatively be measured from low redshift probes of weak lensing, cluster counts or redshift space distortion. We do not intend to provide a complete compilation of all the reported measurements from these probes and we will refer to Figure~4 in \cite{Abdalla:2022yfr} instead (that corresponds to an update of Table II in \cite{Perivolaropoulos:2021jda}). Although the level of tension between the low redshift probes and CMB values ranges from agreement at 1$\sigma$ to a $\sim 3\sigma$ tension, there is a clear trend for low redshift probes to prefer lower values than Planck2018. Since the dark interacting scenario is more amicable to lower values of $\sigma_8$ (or $S_8$) than $\Lambda$CDM, it is opportune to examine the impact of adding these data to our baseline dataset.

The incorporation of the low redshift information will be carried out in a simplified manner that consists in adding to our baseline the {\it measured} values of the parameter $S_8$ as a Gaussian likelihood. There are several values for this parameter available in the literature and the obtained constraints will slightly depend on the concrete value that we incorporate. However, our interest is not to obtain the most precise constraint on the interacting parameter, but to analyse the impact of adding data that are in tension with Planck within $\Lambda$CDM. With this purpose in mind, it is not particularly relevant which value we take. A cautionary comment is however in order because the values of $S_8$ are routinely obtained for $\Lambda$CDM so its use for a different model is not justified a priori. However, we insist that our aim here is to study the impact of generically adding data sensitive to the low-redshift clustering in a simplified manner. Furthermore, since our scenario does not represent a dramatic change with respect to $\Lambda$CDM, we expect our approach to give a good first intuition on the impact of these data for the interaction. For a more detailed discussion on this issue see \cite{BeltranJimenez:2021wbq}.

The results for the baseline$+S_8$ data from Planck SZ cluster counts obtained in \cite{Figueruelo:2021elm} are shown in the red contours in Figure~\ref{Fig:S8-vs-alpha}. The inclusion of the $S_8$ data has a major impact on the results: not only does the fit substantially improve with respect to $\Lambda$CDM, but, more interestingly, the interaction can be properly constrained! Both of these effects are driven by the lower values of $\sigma_8$ that can be accommodated thanks to the clustering suppression produced by the interaction. It is remarkable the improvement in the fit with a $\Delta \chi^2\simeq 24$ \cite{Figueruelo:2021elm}, but it is more remarkable to corroborate that the Gaussian likelihood from $S_8$ selects a preferred non-vanishing value of the interaction parameter $\alpha=1.005^{+0.26}_{-0.33}$, i.e., we find a {\it detection} of the interaction at more than $3\sigma$'s.

\begin{figure}[t]
\includegraphics[width=0.325\textwidth]{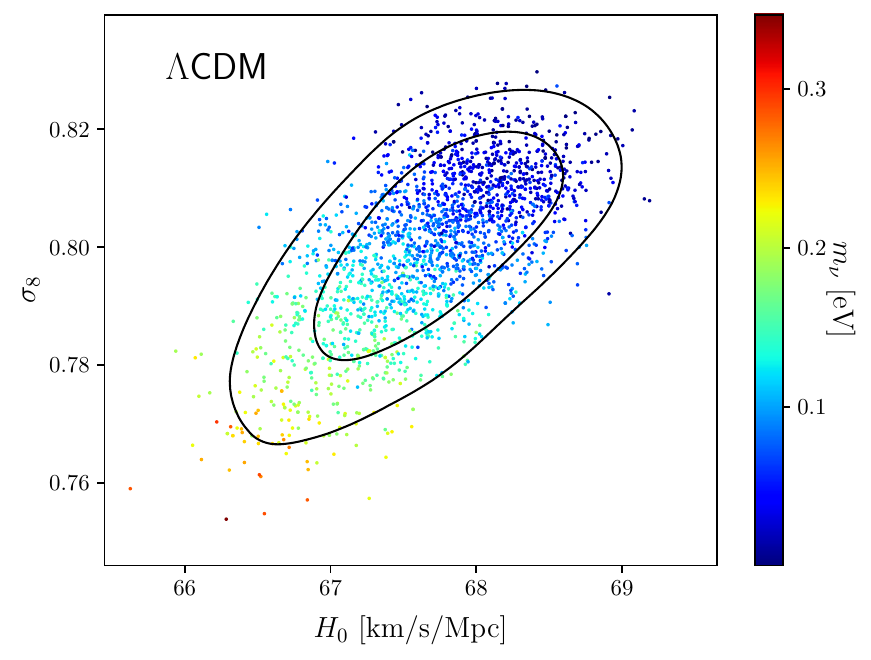}
\includegraphics[width=0.325\textwidth]{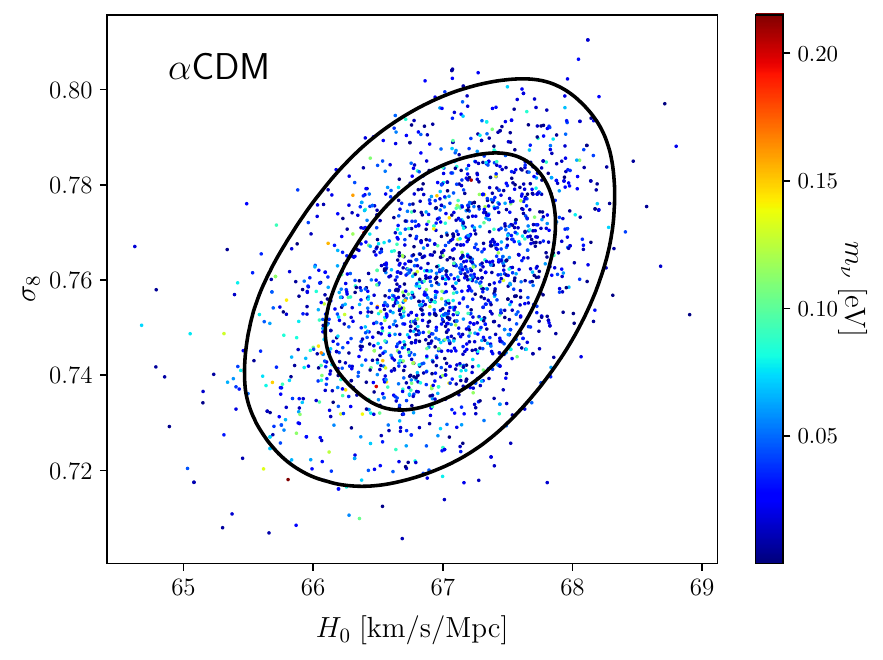}
\includegraphics[width=0.325\textwidth]{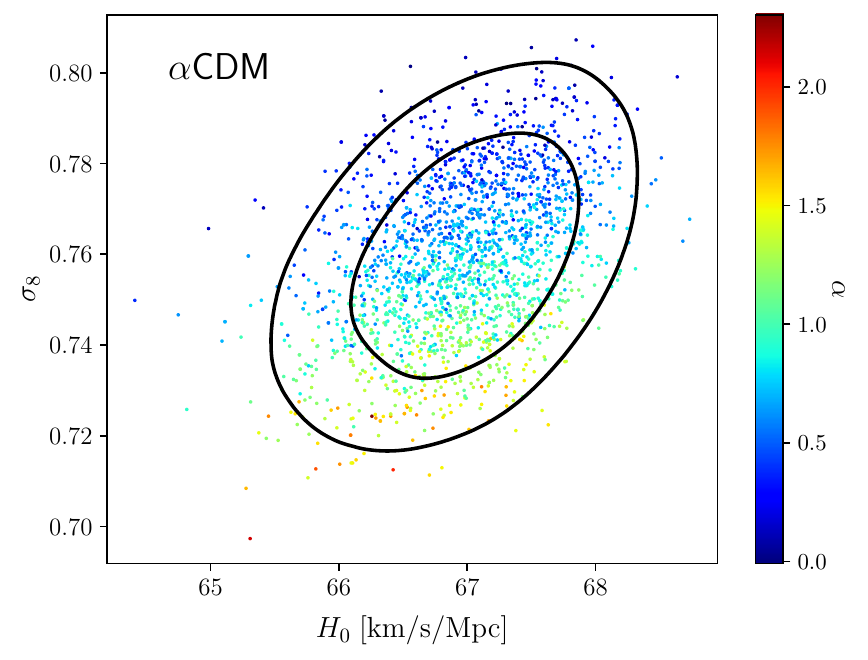}
\caption{Constraints in the $\sigma_8-H_0$ plane for $\Lambda$CDM (left) and $\alpha$CDM (middle and right) allowing for varying neutrino mass and using the baseline+$S_8$ dataset with a Gaussian likelihood corresponding to the DES survey third year release $S_8^{\rm DES-Y3}=0.776\pm0.017$~\cite{DES:2021wwk}. The right panel clearly shows, for $\Lambda$CDM, the correlation between more massive neutrinos and lower values of $\sigma_8$, but at the expense of also obtaining lower values of $H_0$. For $\alpha$CDM however, we see that both correlations disappear (middle panel) and it is $\alpha$ what entirely drives the possibility of having lower values of $\sigma_8$ without any effect on $H_0$ (right panel).}
\label{Fig:S8-H0-vs-mnu_with_low_z}
\end{figure}

We have already mentioned that the inclusion of $S_8$ as a Gaussian likelihood has some caveats and the robustness of the obtained results need to be scrutinise more thoroughly. In this respect, the suppression of the matter power spectrum caused by the interaction could remind to the effects of having massive neutrinos, which produce a similar suppression. A natural question would then be whether allowing for massive neutrinos could jeopardise our findings. This issue has been investigated in \cite{BeltranJimenez:2024lml} where it has been confirmed that the neutrino mass does not degrade the detection of $\alpha$. In fact, although the massive neutrinos can also reduce the clustering, this is another example where ameliorating the $\sigma_8$ tension comes hand-in-hand with an aggravation of the $H_0$ tension (see Figure~\ref{Fig:S8-H0-vs-mnu_with_low_z}). When the interaction is introduced, the massive neutrinos have a very marginal effect. The reason for this is that massive neutrinos do not only suppress the growth of structures, but they also affect the CMB and the background evolution so they are more restricted to efficiently erase structures. We find here one of the advantages of the elastic interacting scenario where the suppression of the clustering is completely decoupled from other effects, in particular, the background evolution and the CMB.

%%%%%%%%%%%%%%%%%%%%%%%%%%%%%%%%%%%%%%%%
\section{Looking ahead: A smoking gun!}
\label{sec:nonlinear}

In view of the exciting results that we obtain when confronting the interacting scenario to current data, it is pertinent to ask how future data will be able to discern the presence of the interaction or to rule it out. Firstly, we can consider future galaxy surveys that will yield tighter constraints on the clustering. This has been carried out in \cite{Figueruelo:2021elm} by performing a forecast analysis for surveys like J-PAS, Euclid and DESI. Assuming a fiducial cosmology with $\alpha=1$, it was found that the specifications of these surveys will constrain the value of $\alpha$ to the percent level with a forecasted error of $\Delta \alpha\simeq 0.06$ and this error is robust against marginalisation of the galaxy bias in each redshift bin. The precision in the measurement of the interaction parameter will be possible because the growth factor acquires a distinctive scale-dependence in the dark-coupling epoch that we will be able to see with those surveys. In fact, the forecasted tight constraint on the interaction parameter is driven by galaxy clustering data, while weak lensing will have a much poorer performance and will not be able to constraint the interaction parameter by itself.

Although constraining the interaction parameter with future data is desirable, a more attractive feature would be to have a unique and definitive prediction of these scenarios. One of such predictions is that the elastic interaction leaves a specific imprint in the dipole of the matter power spectrum that can eventually be detected with SKA-like surveys \cite{Braun:2015B3}. This very distinctive signature can be understood from the Euler equation for the dark matter component \eqref{eq:thetaDM}, which can be simplified for small (but still linear) scales where the peculiar velocities of dark energy are much smaller than the dark matter velocities. In that case, the Euler equation for dark matter can be written as
\begin{equation}
\vec{v}^{\: \prime}_{\rm dm}+\mathcal{H}\Big(1+\Theta\Big) \vec{v}_{\rm dm} +\nabla \Phi=0\,, 
\end{equation}
where we have introduced 
\begin{equation}
\Theta\equiv\alpha\frac{H_0 a}{\Omega_{\text{dm}}(a)\mathcal{H}}\,.
\end{equation}
In this regime, the interaction effectively modifies the friction term of the Euler equation. This modification is analogous to the effect studied in \cite{Bonvin:2018ckp} as a test of the equivalence principle and where it was shown that future SKA data could have enough sensitivity to measure it provided it is above a certain threshold. A similar analysis has been performed in \cite{BeltranJimenez:2022irm} for the dark elastic coupling and it has been found that, if the elastic interaction is to be a resolution for the lack of clustering at low redshift, the effect on the galaxy power spectrum dipole serves as a smoking gun because we must observe it and with the correct amplitude. It has been reported in \cite{BeltranJimenez:2022irm} that the specifications of an SKA-like experiment should have sufficient sensitivity to detect $\alpha$. This represents a hallmark of the dark interaction due to the simplicity of having one single extra parameter to describe the interaction and with no room to hide.

\begin{figure}[!]
\centering
\includegraphics[width=0.9\textwidth]{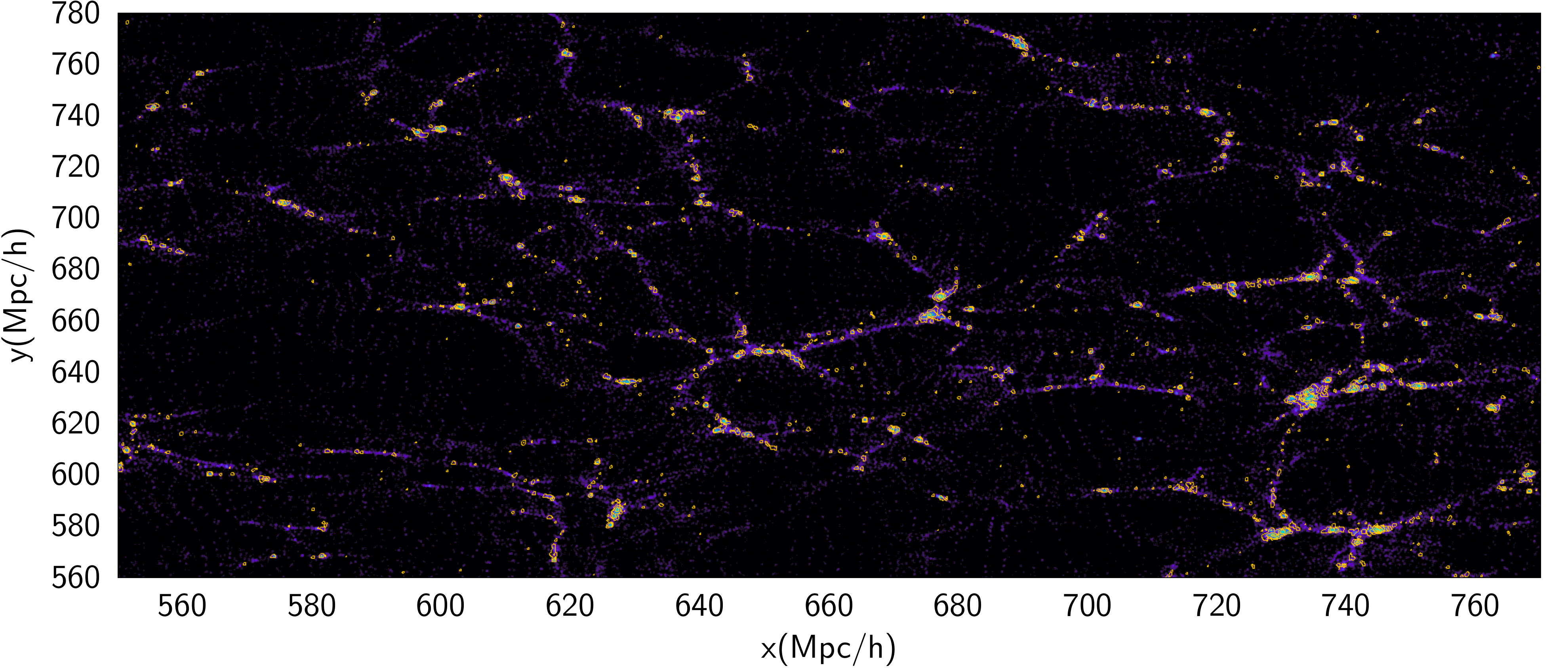}
\includegraphics[width=0.9\textwidth]{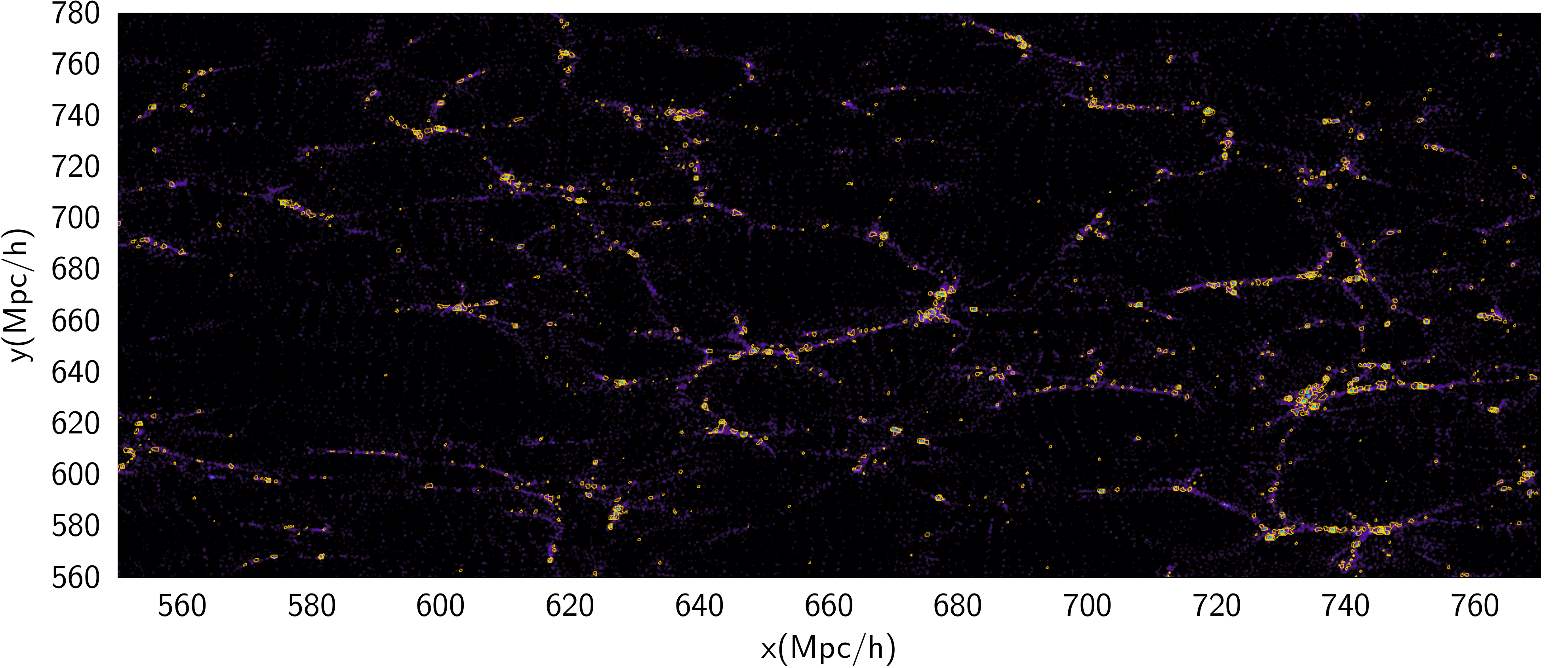}
\caption{Distribution of matter for $\Lambda$CDM (upper) and the $\alpha$CDM (bottom). Dark matter is represented by the black-green density map while regions of high density of baryons are represented by yellow density isolines. We confirm that the interaction gives rise to a less clustered universe and that baryons are locked inside the dark matter haloes.}
\label{fig:nbody_densitymap}
\end{figure}

\begin{figure}[h!]
\centering
\includegraphics[width=0.45\textwidth]{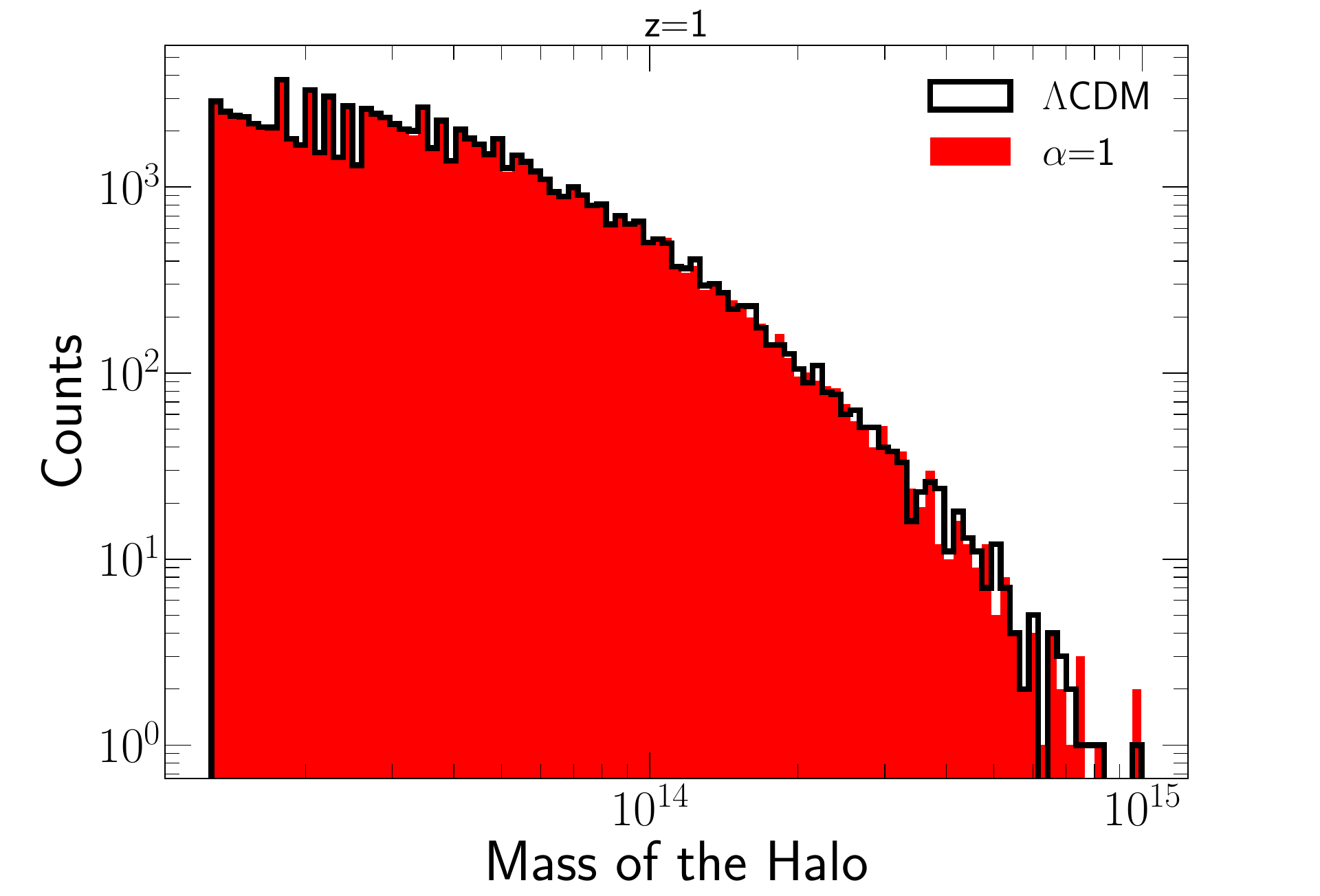} \hspace{0.1cm}
\includegraphics[width=0.45\textwidth]{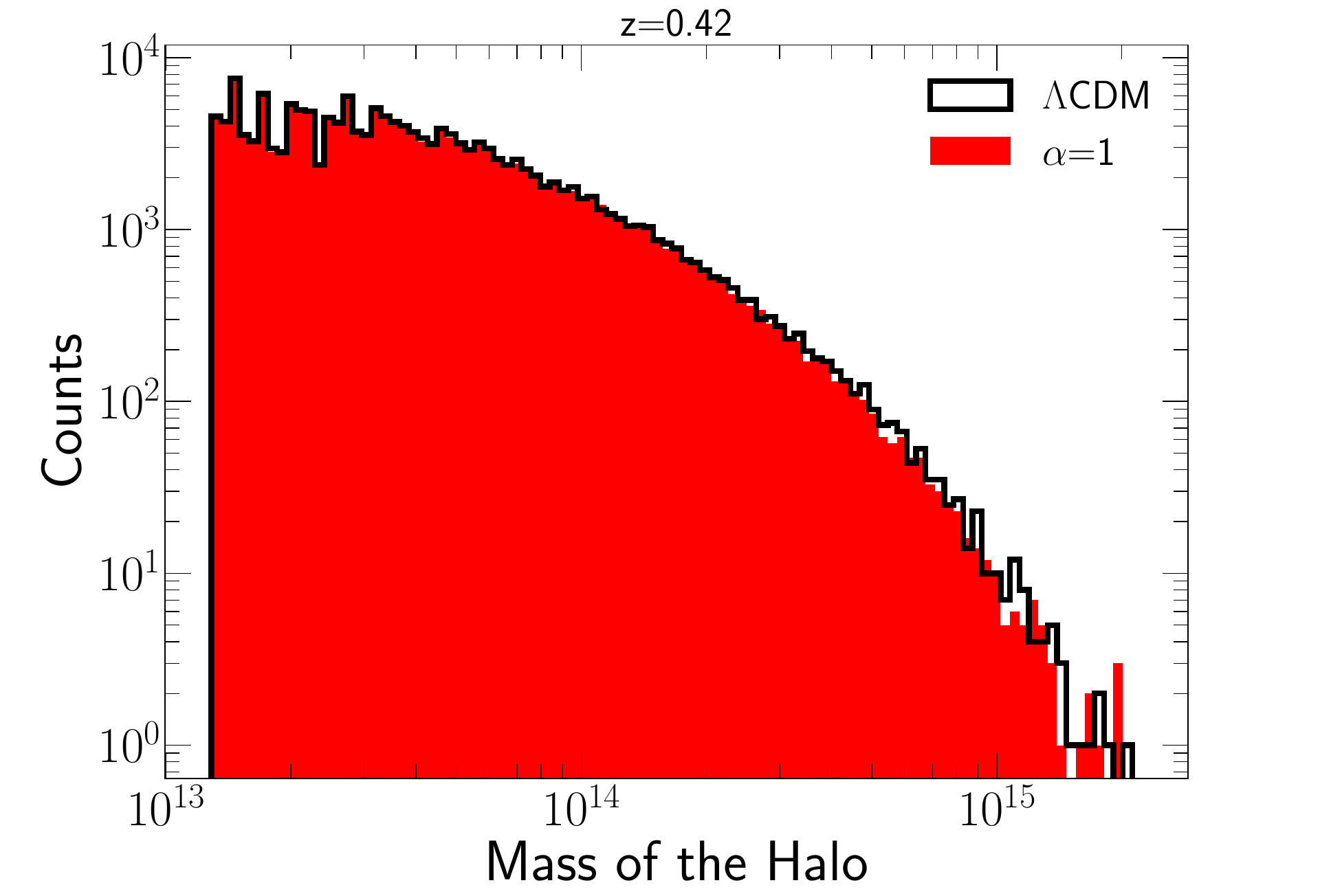}  \\
\includegraphics[width=0.45\textwidth]{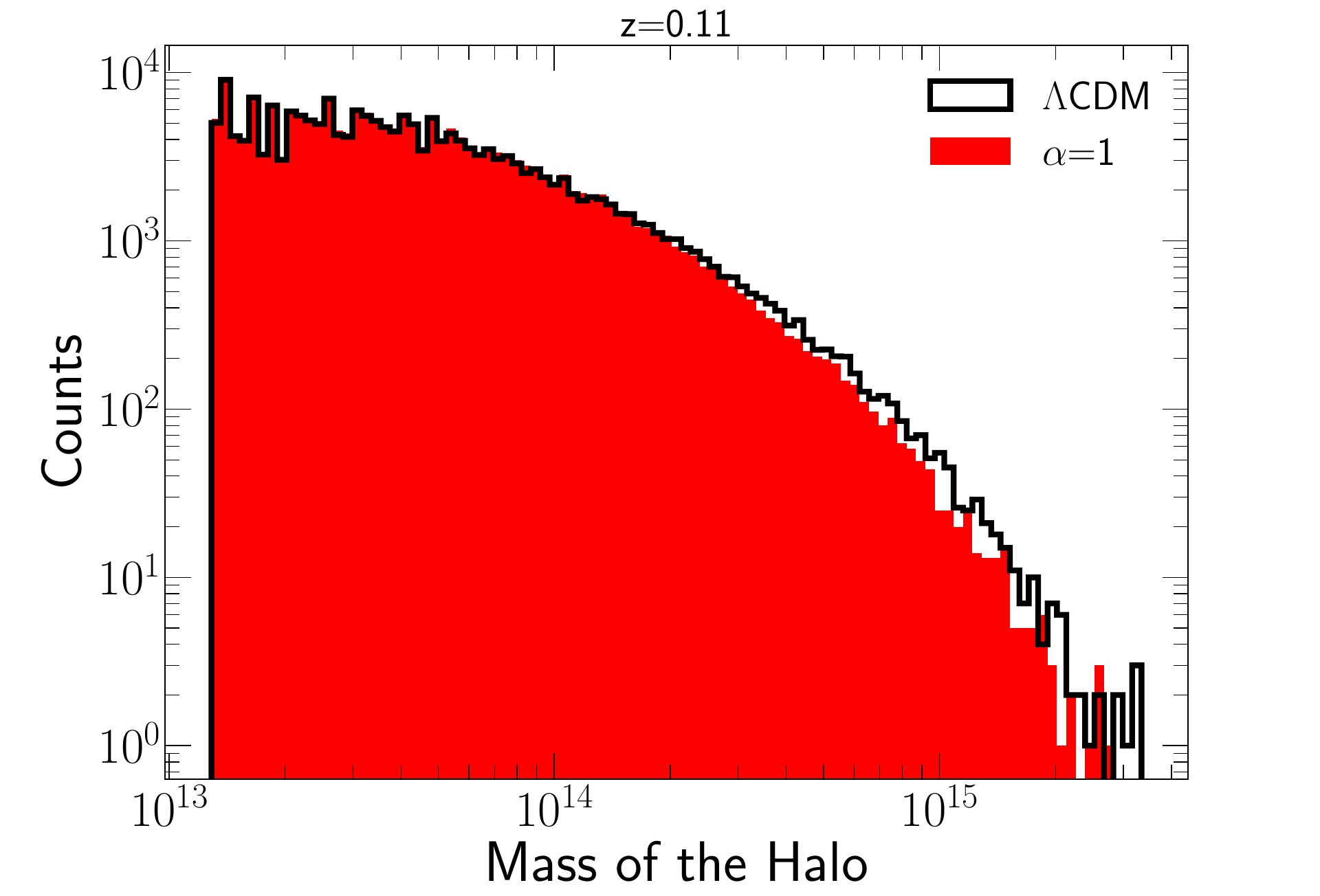} \hspace{0.1cm}
\includegraphics[width=0.45\textwidth]{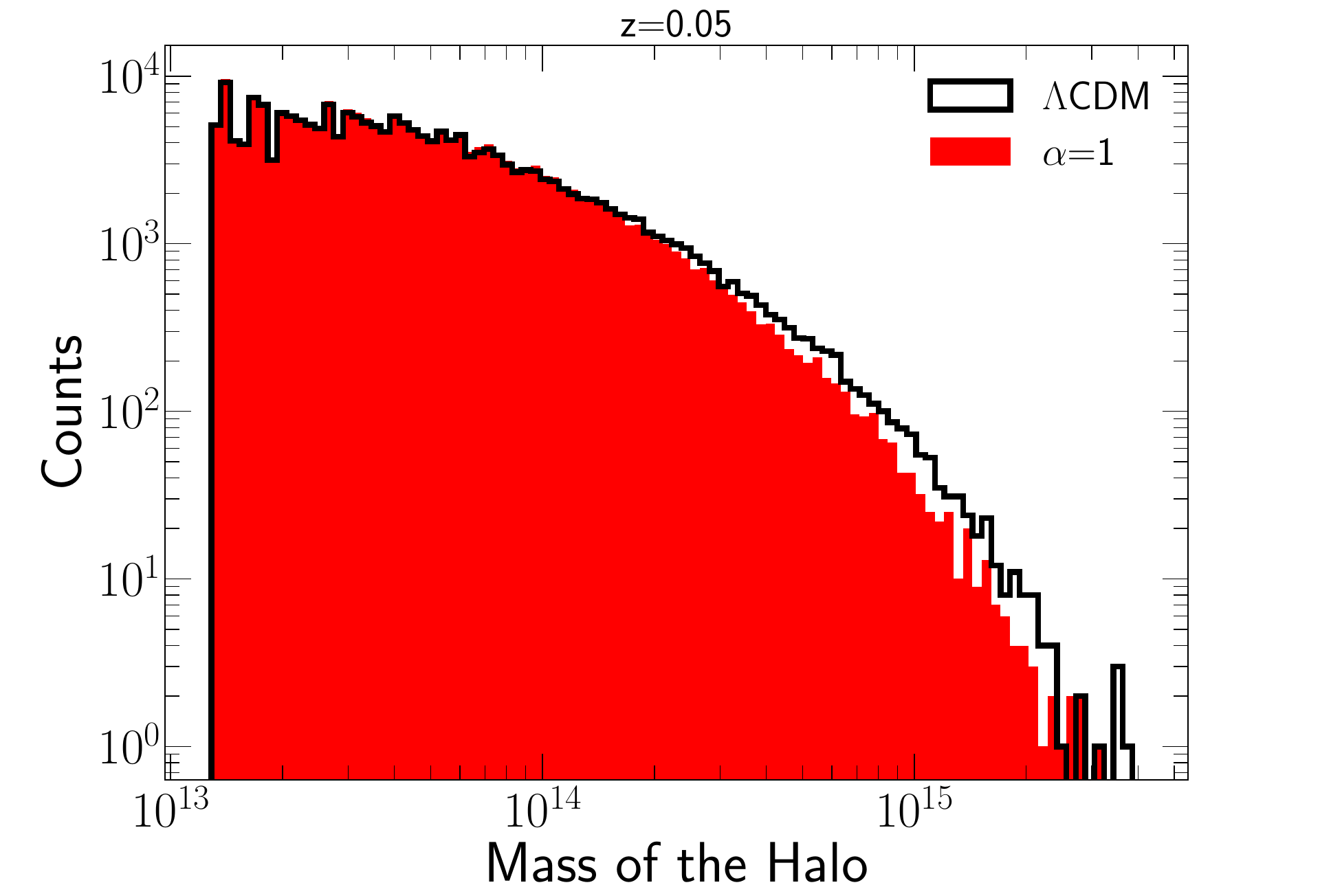}
\caption{Distribution of halos according to their mass in $M_{\odot}/h$ units for $\Lambda$CDM and $\alpha$CDM at different redshifts. These histograms show how the interaction leads to fewer massive haloes as compared to $\Lambda$CDM.}
\label{fig:nbody_histo}
\end{figure}

One could complain that the elastic interaction with dark energy is felt only by dark matter, while the baryons that we actually observe in galaxies are oblivious to it. Thus, the question is whether galaxies are good tracers of the dark matter velocity field or rather, they probe the underlying gravitational potential. Intuitively, one can argue that galaxies are mostly formed at sufficiently high redshift when the interaction does not play any role. Once the galaxy has formed and virialises together with the dark matter halo, it will be attached to it. This means that galaxies will follow the dark matter haloes, thus being good tracers of the dark matter velocity field and the interaction is not expected to be strong enough to disrupt the galaxy-halo system. Of course, this intuition should be confirmed by a proper $N$-body simulation. To test this intuition, we have modified the RAMSES code to include the effects of the interaction \cite{elasticNBody,figueruelo-hernan-2023}. In order to distinguish between baryons and dark matter particles, we have implemented two matter species and only for one of them we have included the interaction with dark energy. In Figure~\ref{fig:nbody_densitymap} we confirm that our intuition is correct and indeed baryons remain attached to the dark matter haloes. As a bonus, in Figure~\ref{fig:nbody_histo} we show how the interacting model predicts a deficit of massive haloes at low redshift. The lack of structures from $N$-body simulations in scenarios with elastic interactions has also been observed in \cite{Baldi:2014ica,Baldi:2016zom,Ferlito:2022mok,Palma:2023ggq}.

%%%%%%%%%%%%%%%%%%%%%%%%%%%%%%%%%%%%
\section{Discussion}
\label{sec:discussion}

A dark-coupling epoch at low redshift with an elastic interaction between dark matter and dark energy does a fantastic job in alleviating cosmological tensions and LSS data even provide evidence for a potential detection of a novel interaction in the dark sector. In this brief review, we have focused on a particular realisation of this scenario, but similar conclusions have been reached with slightly different models featuring a momentum transfer
(see e.g. \cite{Pourtsidou:2016ico,BeltranJimenez:2021wbq,Carrilho:2022mon,Linton:2021cgd,Liu:2023mwx}) so our findings should be regarded as generic features of these scenarios. However, we should bear in mind that {\it extraordinary claims require extraordinary evidence}. Claiming the discovery of a new type of interaction in the dark sector could certainly be catalogued as extraordinary, but the presented evidence falls short and possess some caveats, so we will abstain from making strong assertions. We will however do claim that these scenarios are worth of further investigations and, for that, we do have presented compelling evidence. Their potential to reconcile cosmological tensions is remarkable and they even provide distinctive signatures that serve as a smoking gun. This would be sufficient to motivate a more in-depth scrutiny. But it is also noteworthy the theoretical simplicity of these scenarios with only one extra parameter and a minimalist modification of the Euler equations. An evident and practical advantage is that modifying existing codes to incorporate these interactions is a far less tedious task than for many other {\it standard modified cosmologies}. It is however noteworthy that, despite their simplicity, they provide novel signatures that can serve to guide the search for cracks in the standard model, ultimately leading to the discovery of new physics and, perhaps, even catalyse the revolutionary phase that cosmology might be currently undergoing and that Stage IV experiments will judge.
\vspace{0.5cm}

\vspace{0.3cm}

%{\footnotesize{ 
{\bf Acknowledgements:} It is a pleasure to thank Miguel Aparicio Resco, Wilmar Cardona, Enea Di Dio, Antonio L. Maroto, David F. Mota, Shinji Tsujikawa and Hans Winther for collaborations on these topics. We also thank Pierre Fleury,  Carlos Garc\'ia Garc\'ia and Sunny Vagnozzi for useful discussions. This work was supported by the Project PID2021-122938NB-I00 funded by the Spanish “Ministerio de Ciencia e Innovaci\'on" and FEDER “A way of making Europe” and the Project SA097P24 funded by Junta de Castilla y Le\'on. DB and JBJ also acknowledge support from project PIC-2022-02 funded by Salamanca University. DB also acknowledge support from project PID2022-139841NB-I00 funded
by the Spanish “Ministerio de Ciencia e Innovación”. DF acknowledges support from the programme {\it Ayudas para Financiar la Contrataci\'on Predoctoral de Personal Investigador (ORDEN EDU/601/2020)} funded by Junta de Castilla y Le\'on and European Social Fund and the financial support provided by the postdoctoral research fellowship NWU PDRF Fund NW.1G01487 by the North-West University and the Centre of Space Research

%} }

\vspace{0.3cm}

{\bf Codes availability:} Modified versions of CAMB and CLASS including the elastic interactions are available upon request.

%\appendix
%\section{Example Appendix Section}
%\label{app1}

%\vspace{-0.2cm}
% \newpage
\bibliographystyle{JHEP}
\bibliography{Refs}

\providecommand{\href}[2]{#2}\begingroup\raggedright\begin{thebibliography}{10}

\bibitem{WMAP:2012nax}
{\scshape WMAP} collaboration, \emph{{Nine-Year Wilkinson Microwave Anisotropy
  Probe (WMAP) Observations: Cosmological Parameter Results}},
  \href{https://doi.org/10.1088/0067-0049/208/2/19}{\emph{Astrophys. J. Suppl.}
  {\bfseries 208} (2013) 19} [\href{https://arxiv.org/abs/1212.5226}{{\ttfamily
  1212.5226}}].

\bibitem{Aghanim:2018eyx}
{\scshape Planck} collaboration, \emph{{Planck 2018 results. VI. Cosmological
  parameters}},
  \href{https://doi.org/10.1051/0004-6361/201833910}{\emph{Astron. Astrophys.}
  {\bfseries 641} (2020) A6}
  [\href{https://arxiv.org/abs/1807.06209}{{\ttfamily 1807.06209}}].

\bibitem{eBOSS:2020yzd}
{\scshape eBOSS} collaboration, \emph{{Completed SDSS-IV extended Baryon
  Oscillation Spectroscopic Survey: Cosmological implications from two decades
  of spectroscopic surveys at the Apache Point Observatory}},
  \href{https://doi.org/10.1103/PhysRevD.103.083533}{\emph{Phys. Rev. D}
  {\bfseries 103} (2021) 083533}
  [\href{https://arxiv.org/abs/2007.08991}{{\ttfamily 2007.08991}}].

\bibitem{DES:2021bvc}
{\scshape DES} collaboration, \emph{{Dark Energy Survey Year 3 results:
  Cosmology from cosmic shear and robustness to data calibration}},
  \href{https://doi.org/10.1103/PhysRevD.105.023514}{\emph{Phys. Rev. D}
  {\bfseries 105} (2022) 023514}
  [\href{https://arxiv.org/abs/2105.13543}{{\ttfamily 2105.13543}}].

\bibitem{DES:2021vln}
{\scshape DES} collaboration, \emph{{Dark Energy Survey Year 3 results:
  Cosmology from cosmic shear and robustness to modeling uncertainty}},
  \href{https://doi.org/10.1103/PhysRevD.105.023515}{\emph{Phys. Rev. D}
  {\bfseries 105} (2022) 023515}
  [\href{https://arxiv.org/abs/2105.13544}{{\ttfamily 2105.13544}}].

\bibitem{DES:2021wwk}
{\scshape DES} collaboration, \emph{{Dark Energy Survey Year 3 results:
  Cosmological constraints from galaxy clustering and weak lensing}},
  \href{https://doi.org/10.1103/PhysRevD.105.023520}{\emph{Phys. Rev. D}
  {\bfseries 105} (2022) 023520}
  [\href{https://arxiv.org/abs/2105.13549}{{\ttfamily 2105.13549}}].

\bibitem{Kuijken:2015vca}
K.~Kuijken et~al., \emph{{Gravitational Lensing Analysis of the Kilo Degree
  Survey}}, \href{https://doi.org/10.1093/mnras/stv2140}{\emph{Mon. Not. Roy.
  Astron. Soc.} {\bfseries 454} (2015) 3500}
  [\href{https://arxiv.org/abs/1507.00738}{{\ttfamily 1507.00738}}].

\bibitem{KiDS:2020suj}
{\scshape KiDS} collaboration, \emph{{KiDS-1000 Cosmology: Cosmic shear
  constraints and comparison between two point statistics}},
  \href{https://doi.org/10.1051/0004-6361/202039070}{\emph{Astron. Astrophys.}
  {\bfseries 645} (2021) A104}
  [\href{https://arxiv.org/abs/2007.15633}{{\ttfamily 2007.15633}}].

\bibitem{Dalal:2023olq}
R.~Dalal et~al., \emph{{Hyper Suprime-Cam Year 3 results: Cosmology from cosmic
  shear power spectra}},
  \href{https://doi.org/10.1103/PhysRevD.108.123519}{\emph{Phys. Rev. D}
  {\bfseries 108} (2023) 123519}
  [\href{https://arxiv.org/abs/2304.00701}{{\ttfamily 2304.00701}}].

\bibitem{Li:2023tui}
X.~Li et~al., \emph{{Hyper Suprime-Cam Year 3 results: Cosmology from cosmic
  shear two-point correlation functions}},
  \href{https://doi.org/10.1103/PhysRevD.108.123518}{\emph{Phys. Rev. D}
  {\bfseries 108} (2023) 123518}
  [\href{https://arxiv.org/abs/2304.00702}{{\ttfamily 2304.00702}}].

\bibitem{DESI:2023ytc}
{\scshape DESI} collaboration, \emph{{The Early Data Release of the Dark Energy
  Spectroscopic Instrument}},
  \href{https://doi.org/10.3847/1538-3881/ad3217}{\emph{Astron. J.} {\bfseries
  168} (2024) 58} [\href{https://arxiv.org/abs/2306.06308}{{\ttfamily
  2306.06308}}].

\bibitem{Perivolaropoulos:2021jda}
L.~Perivolaropoulos and F.~Skara, \emph{{Challenges for
  \ensuremath{\Lambda}CDM: An update}},
  \href{https://doi.org/10.1016/j.newar.2022.101659}{\emph{New Astron. Rev.}
  {\bfseries 95} (2022) 101659}
  [\href{https://arxiv.org/abs/2105.05208}{{\ttfamily 2105.05208}}].

\bibitem{Abdalla:2022yfr}
E.~Abdalla et~al., \emph{{Cosmology intertwined: A review of the particle
  physics, astrophysics, and cosmology associated with the cosmological
  tensions and anomalies}},
  \href{https://doi.org/10.1016/j.jheap.2022.04.002}{\emph{JHEAp} {\bfseries
  34} (2022) 49} [\href{https://arxiv.org/abs/2203.06142}{{\ttfamily
  2203.06142}}].

\bibitem{DESI:2024uvr}
{\scshape DESI} collaboration, \emph{{DESI 2024 III: Baryon Acoustic
  Oscillations from Galaxies and Quasars}},
  \href{https://arxiv.org/abs/2404.03000}{{\ttfamily 2404.03000}}.

\bibitem{DESI:2024mwx}
{\scshape DESI} collaboration, \emph{{DESI 2024 VI: Cosmological Constraints
  from the Measurements of Baryon Acoustic Oscillations}},
  \href{https://arxiv.org/abs/2404.03002}{{\ttfamily 2404.03002}}.

\bibitem{Sakr:2023hrl}
Z.~Sakr, \emph{{Testing the hypothesis of a matter density discrepancy within
  LCDM model using multiple probes}},
  \href{https://doi.org/10.1103/PhysRevD.108.083519}{\emph{Phys. Rev. D}
  {\bfseries 108} (2023) 083519}
  [\href{https://arxiv.org/abs/2305.02846}{{\ttfamily 2305.02846}}].

\bibitem{Poulin:2024ken}
V.~Poulin, T.L.~Smith, R.~Calder\'on and T.~Simon, \emph{{On the implications
  of the `cosmic calibration tension' beyond $H_0$ and the synergy between
  early- and late-time new physics}},
  \href{https://arxiv.org/abs/2407.18292}{{\ttfamily 2407.18292}}.

\bibitem{Pedrotti:2024kpn}
D.~Pedrotti, J.-Q.~Jiang, L.A.~Escamilla, S.S.~da~Costa and S.~Vagnozzi,
  \emph{{Multidimensionality of the Hubble tension: the roles of $\Omega_m$ and
  $\omega_c$}},  \href{https://arxiv.org/abs/2408.04530}{{\ttfamily
  2408.04530}}.

\bibitem{ACT:2023kun}
{\scshape ACT} collaboration, \emph{{The Atacama Cosmology Telescope: DR6
  Gravitational Lensing Map and Cosmological Parameters}},
  \href{https://doi.org/10.3847/1538-4357/acff5f}{\emph{Astrophys. J.}
  {\bfseries 962} (2024) 113}
  [\href{https://arxiv.org/abs/2304.05203}{{\ttfamily 2304.05203}}].

\bibitem{Wang:2024vmw}
B.~Wang, E.~Abdalla, F.~Atrio-Barandela and D.~Pav\'on, \emph{{Further
  understanding the interaction between dark energy and dark matter: current
  status and future directions}},
  \href{https://doi.org/10.1088/1361-6633/ad2527}{\emph{Rept. Prog. Phys.}
  {\bfseries 87} (2024) 036901}
  [\href{https://arxiv.org/abs/2402.00819}{{\ttfamily 2402.00819}}].

\bibitem{Simpson:2010vh}
F.~Simpson, \emph{{Scattering of dark matter and dark energy}},
  \href{https://doi.org/10.1103/PhysRevD.82.083505}{\emph{Phys. Rev. D}
  {\bfseries 82} (2010) 083505}
  [\href{https://arxiv.org/abs/1007.1034}{{\ttfamily 1007.1034}}].

\bibitem{Pourtsidou:2013nha}
A.~Pourtsidou, C.~Skordis and E.J.~Copeland, \emph{{Models of dark matter
  coupled to dark energy}},
  \href{https://doi.org/10.1103/PhysRevD.88.083505}{\emph{Phys. Rev. D}
  {\bfseries 88} (2013) 083505}
  [\href{https://arxiv.org/abs/1307.0458}{{\ttfamily 1307.0458}}].

\bibitem{Skordis:2015yra}
C.~Skordis, A.~Pourtsidou and E.J.~Copeland, \emph{{Parametrized
  post-Friedmannian framework for interacting dark energy theories}},
  \href{https://doi.org/10.1103/PhysRevD.91.083537}{\emph{Phys. Rev. D}
  {\bfseries 91} (2015) 083537}
  [\href{https://arxiv.org/abs/1502.07297}{{\ttfamily 1502.07297}}].

\bibitem{Koivisto:2015qua}
T.S.~Koivisto, E.N.~Saridakis and N.~Tamanini, \emph{{Scalar-Fluid theories:
  cosmological perturbations and large-scale structure}},
  \href{https://doi.org/10.1088/1475-7516/2015/09/047}{\emph{JCAP} {\bfseries
  09} (2015) 047} [\href{https://arxiv.org/abs/1505.07556}{{\ttfamily
  1505.07556}}].

\bibitem{Kase:2019mox}
R.~Kase and S.~Tsujikawa, \emph{{Weak cosmic growth in coupled dark energy with
  a Lagrangian formulation}},
  \href{https://doi.org/10.1016/j.physletb.2020.135400}{\emph{Phys. Lett. B}
  {\bfseries 804} (2020) 135400}
  [\href{https://arxiv.org/abs/1911.02179}{{\ttfamily 1911.02179}}].

\bibitem{Chamings:2019kcl}
F.N.~Chamings, A.~Avgoustidis, E.J.~Copeland, A.M.~Green and A.~Pourtsidou,
  \emph{{Understanding the suppression of structure formation from dark
  matter-dark energy momentum coupling}},
  \href{https://doi.org/10.1103/PhysRevD.101.043531}{\emph{Phys. Rev. D}
  {\bfseries 101} (2020) 043531}
  [\href{https://arxiv.org/abs/1912.09858}{{\ttfamily 1912.09858}}].

\bibitem{Amendola:2020ldb}
L.~Amendola and S.~Tsujikawa, \emph{{Scaling solutions and weak gravity in dark
  energy with energy and momentum couplings}},
  \href{https://doi.org/10.1088/1475-7516/2020/06/020}{\emph{JCAP} {\bfseries
  06} (2020) 020} [\href{https://arxiv.org/abs/2003.02686}{{\ttfamily
  2003.02686}}].

\bibitem{Linton:2021cgd}
M.S.~Linton, R.~Crittenden and A.~Pourtsidou, \emph{{Momentum transfer models
  of interacting dark energy}},
  \href{https://doi.org/10.1088/1475-7516/2022/08/075}{\emph{JCAP} {\bfseries
  08} (2022) 075} [\href{https://arxiv.org/abs/2107.03235}{{\ttfamily
  2107.03235}}].

\bibitem{ManciniSpurio:2021jvx}
A.~Mancini~Spurio and A.~Pourtsidou, \emph{{KiDS-1000 cosmology: machine
  learning \textendash{} accelerated constraints on interacting dark energy
  with CosmoPower}}, \href{https://doi.org/10.1093/mnrasl/slac019}{\emph{Mon.
  Not. Roy. Astron. Soc.} {\bfseries 512} (2022) L44}
  [\href{https://arxiv.org/abs/2110.07587}{{\ttfamily 2110.07587}}].

\bibitem{Nakamura:2019phn}
S.~Nakamura, R.~Kase and S.~Tsujikawa, \emph{{Coupled vector dark energy}},
  \href{https://doi.org/10.1088/1475-7516/2019/12/032}{\emph{JCAP} {\bfseries
  12} (2019) 032} [\href{https://arxiv.org/abs/1907.12216}{{\ttfamily
  1907.12216}}].

\bibitem{DeFelice:2020icf}
A.~De~Felice, S.~Nakamura and S.~Tsujikawa, \emph{{Suppressed cosmic growth in
  coupled vector-tensor theories}},
  \href{https://doi.org/10.1103/PhysRevD.102.063531}{\emph{Phys. Rev. D}
  {\bfseries 102} (2020) 063531}
  [\href{https://arxiv.org/abs/2004.09384}{{\ttfamily 2004.09384}}].

\bibitem{Cardona:2023gzq}
W.~Cardona, J.L.~Palacios-C\'ordoba and C.A.~Valenzuela-Toledo,
  \emph{{Scrutinizing coupled vector dark energy in light of data}},
  \href{https://doi.org/10.1088/1475-7516/2024/04/016}{\emph{JCAP} {\bfseries
  04} (2024) 016} [\href{https://arxiv.org/abs/2310.13877}{{\ttfamily
  2310.13877}}].

\bibitem{Pookkillath:2024ycd}
M.C.~Pookkillath and K.~Koyama, \emph{{Theory of interacting vector dark energy
  and fluid}},  \href{https://arxiv.org/abs/2405.06565}{{\ttfamily
  2405.06565}}.

\bibitem{Gleyzes:2015pma}
J.~Gleyzes, D.~Langlois, M.~Mancarella and F.~Vernizzi, \emph{{Effective Theory
  of Interacting Dark Energy}},
  \href{https://doi.org/10.1088/1475-7516/2015/08/054}{\emph{JCAP} {\bfseries
  08} (2015) 054} [\href{https://arxiv.org/abs/1504.05481}{{\ttfamily
  1504.05481}}].

\bibitem{Asghari:2019qld}
M.~Asghari, J.~Beltr\'an~Jim\'enez, S.~Khosravi and D.F.~Mota, \emph{{On
  structure formation from a small-scales-interacting dark sector}},
  \href{https://doi.org/10.1088/1475-7516/2019/04/042}{\emph{JCAP} {\bfseries
  04} (2019) 042} [\href{https://arxiv.org/abs/1902.05532}{{\ttfamily
  1902.05532}}].

\bibitem{BeltranJimenez:2020qdu}
J.~Beltr\'an~Jim\'enez, D.~Bettoni, D.~Figueruelo, F.A.~Teppa~Pannia and
  S.~Tsujikawa, \emph{{Velocity-dependent interacting dark energy and dark
  matter with a Lagrangian description of perfect fluids}},
  \href{https://doi.org/10.1088/1475-7516/2021/03/085}{\emph{JCAP} {\bfseries
  03} (2021) 085} [\href{https://arxiv.org/abs/2012.12204}{{\ttfamily
  2012.12204}}].

\bibitem{Vagnozzi:2019kvw}
S.~Vagnozzi, L.~Visinelli, O.~Mena and D.F.~Mota, \emph{{Do we have any hope of
  detecting scattering between dark energy and baryons through cosmology?}},
  \href{https://doi.org/10.1093/mnras/staa311}{\emph{Mon. Not. Roy. Astron.
  Soc.} {\bfseries 493} (2020) 1139}
  [\href{https://arxiv.org/abs/1911.12374}{{\ttfamily 1911.12374}}].

\bibitem{BeltranJimenez:2020iyx}
J.~Beltr\'an~Jim\'enez, D.~Bettoni, D.~Figueruelo and F.A.~Teppa~Pannia,
  \emph{{On cosmological signatures of baryons-dark energy elastic couplings}},
  \href{https://doi.org/10.1088/1475-7516/2020/08/020}{\emph{JCAP} {\bfseries
  08} (2020) 020} [\href{https://arxiv.org/abs/2004.14661}{{\ttfamily
  2004.14661}}].

\bibitem{Wilkinson:2013kia}
R.J.~Wilkinson, J.~Lesgourgues and C.~Boehm, \emph{{Using the CMB angular power
  spectrum to study Dark Matter-photon interactions}},
  \href{https://doi.org/10.1088/1475-7516/2014/04/026}{\emph{JCAP} {\bfseries
  04} (2014) 026} [\href{https://arxiv.org/abs/1309.7588}{{\ttfamily
  1309.7588}}].

\bibitem{Stadler:2018jin}
J.~Stadler and C.~B\oe{}hm, \emph{{Constraints on $\gamma$-CDM interactions
  matching the Planck data precision}},
  \href{https://doi.org/10.1088/1475-7516/2018/10/009}{\emph{JCAP} {\bfseries
  10} (2018) 009} [\href{https://arxiv.org/abs/1802.06589}{{\ttfamily
  1802.06589}}].

\bibitem{Kumar:2018yhh}
S.~Kumar, R.C.~Nunes and S.K.~Yadav, \emph{{Cosmological bounds on dark
  matter-photon coupling}},
  \href{https://doi.org/10.1103/PhysRevD.98.043521}{\emph{Phys. Rev. D}
  {\bfseries 98} (2018) 043521}
  [\href{https://arxiv.org/abs/1803.10229}{{\ttfamily 1803.10229}}].

\bibitem{Stadler:2019dii}
J.~Stadler, C.~B\oe{}hm and O.~Mena, \emph{{Comprehensive Study of
  Neutrino-Dark Matter Mixed Damping}},
  \href{https://doi.org/10.1088/1475-7516/2019/08/014}{\emph{JCAP} {\bfseries
  08} (2019) 014} [\href{https://arxiv.org/abs/1903.00540}{{\ttfamily
  1903.00540}}].

\bibitem{Figueruelo:2021elm}
D.~Figueruelo et~al., \emph{{J-PAS: Forecasts for dark matter - dark energy
  elastic couplings}},
  \href{https://doi.org/10.1088/1475-7516/2021/07/022}{\emph{JCAP} {\bfseries
  07} (2021) 022} [\href{https://arxiv.org/abs/2103.01571}{{\ttfamily
  2103.01571}}].

\bibitem{BeltranJimenez:2021wbq}
J.~Beltr\'an~Jim\'enez, D.~Bettoni, D.~Figueruelo, F.A.~Teppa~Pannia and
  S.~Tsujikawa, \emph{{Probing elastic interactions in the dark sector and the
  role of S8}}, \href{https://doi.org/10.1103/PhysRevD.104.103503}{\emph{Phys.
  Rev. D} {\bfseries 104} (2021) 103503}
  [\href{https://arxiv.org/abs/2106.11222}{{\ttfamily 2106.11222}}].

\bibitem{Cardona:2022mdq}
W.~Cardona and D.~Figueruelo, \emph{{Momentum transfer in the dark sector and
  lensing convergence in upcoming galaxy surveys}},
  \href{https://doi.org/10.1088/1475-7516/2022/12/010}{\emph{JCAP} {\bfseries
  12} (2022) 010} [\href{https://arxiv.org/abs/2209.12583}{{\ttfamily
  2209.12583}}].

\bibitem{Poulin:2022sgp}
V.~Poulin, J.L.~Bernal, E.D.~Kovetz and M.~Kamionkowski, \emph{{Sigma-8 tension
  is a drag}}, \href{https://doi.org/10.1103/PhysRevD.107.123538}{\emph{Phys.
  Rev. D} {\bfseries 107} (2023) 123538}
  [\href{https://arxiv.org/abs/2209.06217}{{\ttfamily 2209.06217}}].

\bibitem{Ma:1995ey}
C.-P.~Ma and E.~Bertschinger, \emph{{Cosmological perturbation theory in the
  synchronous and conformal Newtonian gauges}},
  \href{https://doi.org/10.1086/176550}{\emph{Astrophys. J.} {\bfseries 455}
  (1995) 7} [\href{https://arxiv.org/abs/astro-ph/9506072}{{\ttfamily
  astro-ph/9506072}}].

\bibitem{weinberg2008cosmology}
S.~Weinberg, \emph{Cosmology}, Cosmology, OUP Oxford (2008).

\bibitem{Kumar:2017bpv}
S.~Kumar and R.C.~Nunes, \emph{{Observational constraints on dark
  matter\textendash{}dark energy scattering cross section}},
  \href{https://doi.org/10.1140/epjc/s10052-017-5334-3}{\emph{Eur. Phys. J. C}
  {\bfseries 77} (2017) 734}
  [\href{https://arxiv.org/abs/1709.02384}{{\ttfamily 1709.02384}}].

\bibitem{Garcia-Garcia:2024gzy}
C.~Garc\'\i{}a-Garc\'\i{}a, M.~Zennaro, G.~Aric\`o, D.~Alonso and R.E.~Angulo,
  \emph{{Cosmic shear with small scales: DES-Y3, KiDS-1000 and HSC-DR1}},
  \href{https://doi.org/10.1088/1475-7516/2024/08/024}{\emph{JCAP} {\bfseries
  08} (2024) 024} [\href{https://arxiv.org/abs/2403.13794}{{\ttfamily
  2403.13794}}].

\bibitem{Jedamzik:2020zmd}
K.~Jedamzik, L.~Pogosian and G.-B.~Zhao, \emph{{Why reducing the cosmic sound
  horizon alone can not fully resolve the Hubble tension}},
  \href{https://doi.org/10.1038/s42005-021-00628-x}{\emph{Commun. in Phys.}
  {\bfseries 4} (2021) 123} [\href{https://arxiv.org/abs/2010.04158}{{\ttfamily
  2010.04158}}].

\bibitem{Sanchez:2020vvb}
A.G.~Sanchez, \emph{{Arguments against using $h^{-1}{\rm Mpc}$ units in
  observational cosmology}},
  \href{https://doi.org/10.1103/PhysRevD.102.123511}{\emph{Phys. Rev. D}
  {\bfseries 102} (2020) 123511}
  [\href{https://arxiv.org/abs/2002.07829}{{\ttfamily 2002.07829}}].

\bibitem{BeltranJimenez:2024lml}
J.~Beltr\'an~Jim\'enez, D.~Figueruelo and F.A.~Teppa~Pannia,
  \emph{{Nondegeneracy of massive neutrinos and elastic interactions in the
  dark sector}}, \href{https://doi.org/10.1103/PhysRevD.110.023527}{\emph{Phys.
  Rev. D} {\bfseries 110} (2024) 023527}
  [\href{https://arxiv.org/abs/2403.03216}{{\ttfamily 2403.03216}}].

\bibitem{Braun:2015B3}
R.~Braun, T.L.~Bourke, J.A.~Green, E.~Keane and J.~Wagg, \emph{{Advancing
  Astrophysics with the Square Kilometre Array}},
  \href{https://doi.org/10.22323/1.215.0174}{\emph{PoS} {\bfseries AASKA14}
  (2015) 174}.

\bibitem{Bonvin:2018ckp}
C.~Bonvin and P.~Fleury, \emph{{Testing the equivalence principle on
  cosmological scales}},
  \href{https://doi.org/10.1088/1475-7516/2018/05/061}{\emph{JCAP} {\bfseries
  05} (2018) 061} [\href{https://arxiv.org/abs/1803.02771}{{\ttfamily
  1803.02771}}].

\bibitem{BeltranJimenez:2022irm}
J.~Beltr\'an~Jim\'enez, E.~Di~Dio and D.~Figueruelo, \emph{{A smoking gun from
  the power spectrum dipole for elastic interactions in the dark sector}},
  \href{https://doi.org/10.1088/1475-7516/2023/11/088}{\emph{JCAP} {\bfseries
  11} (2023) 088} [\href{https://arxiv.org/abs/2212.08617}{{\ttfamily
  2212.08617}}].

\bibitem{elasticNBody}
J.~Beltr\'an~Jim\'enez, D.~Figueruelo, D.F.~Mota and H.~Winther,
  \emph{{Non-linear structure formation in the presence of elastic dark
  interactions}}, {\emph{to appear} (2024) }.

\bibitem{figueruelo-hernan-2023}
D.~Figueruelo~Hern\'an, \emph{{Restrictions in the dark sector of the universe
  and modified gravity with large scale structure and gravitational waves}},
  Ph.D. thesis, Universidad de Salamanca, 2023.

\bibitem{Baldi:2014ica}
M.~Baldi and F.~Simpson, \emph{{Simulating Momentum Exchange in the Dark
  Sector}}, \href{https://doi.org/10.1093/mnras/stv405}{\emph{Mon. Not. Roy.
  Astron. Soc.} {\bfseries 449} (2015) 2239}
  [\href{https://arxiv.org/abs/1412.1080}{{\ttfamily 1412.1080}}].

\bibitem{Baldi:2016zom}
M.~Baldi and F.~Simpson, \emph{{Structure formation simulations with momentum
  exchange: alleviating tensions between high-redshift and low-redshift
  cosmological probes}},
  \href{https://doi.org/10.1093/mnras/stw2702}{\emph{Mon. Not. Roy. Astron.
  Soc.} {\bfseries 465} (2017) 653}
  [\href{https://arxiv.org/abs/1605.05623}{{\ttfamily 1605.05623}}].

\bibitem{Ferlito:2022mok}
F.~Ferlito, S.~Vagnozzi, D.F.~Mota and M.~Baldi, \emph{{Cosmological direct
  detection of dark energy: Non-linear structure formation signatures of dark
  energy scattering with visible matter}},
  \href{https://doi.org/10.1093/mnras/stac649}{\emph{Mon. Not. Roy. Astron.
  Soc.} {\bfseries 512} (2022) 1885}
  [\href{https://arxiv.org/abs/2201.04528}{{\ttfamily 2201.04528}}].

\bibitem{Palma:2023ggq}
D.~Palma and G.N.~Candlish, \emph{{Cosmological simulations of a momentum
  coupling between dark matter and quintessence}},
  \href{https://doi.org/10.1093/mnras/stad2739}{\emph{Mon. Not. Roy. Astron.
  Soc.} {\bfseries 526} (2023) 1904}
  [\href{https://arxiv.org/abs/2309.04530}{{\ttfamily 2309.04530}}].

\bibitem{Pourtsidou:2016ico}
A.~Pourtsidou and T.~Tram, \emph{{Reconciling CMB and structure growth
  measurements with dark energy interactions}},
  \href{https://doi.org/10.1103/PhysRevD.94.043518}{\emph{Phys. Rev. D}
  {\bfseries 94} (2016) 043518}
  [\href{https://arxiv.org/abs/1604.04222}{{\ttfamily 1604.04222}}].

\bibitem{Carrilho:2022mon}
P.~Carrilho, C.~Moretti and A.~Pourtsidou, \emph{{Cosmology with the EFTofLSS
  and BOSS: dark energy constraints and a note on priors}},
  \href{https://doi.org/10.1088/1475-7516/2023/01/028}{\emph{JCAP} {\bfseries
  01} (2023) 028} [\href{https://arxiv.org/abs/2207.14784}{{\ttfamily
  2207.14784}}].

\bibitem{Liu:2023mwx}
X.~Liu, S.~Tsujikawa and K.~Ichiki, \emph{{Observational constraints on
  interactions between dark energy and dark matter with momentum and energy
  transfers}}, \href{https://doi.org/10.1103/PhysRevD.109.043533}{\emph{Phys.
  Rev. D} {\bfseries 109} (2024) 043533}
  [\href{https://arxiv.org/abs/2309.13946}{{\ttfamily 2309.13946}}].

\end{thebibliography}\endgroup

\end{document}